\documentclass[sigconf, screen]{acmart}

\makeatletter
\input{dvipsnam.def}
\makeatother

\usepackage{algorithm}
\usepackage{algorithmic}
\usepackage{textcomp}
\usepackage{listings}
\usepackage{pifont}
\usepackage{multirow}
\usepackage{enumitem}

\newcommand{\name}{AIM}
\newcommand{\lhr}{LHR}

\newcommand{\tog}{\text{R}\textsubscript{tog}}
\newcommand{\HR}{HR}

\newcommand{\WDS}{{Weight Distribution Shift}}
\newcommand{\wds}{{WDS}}
\newcommand{\booster}{{IR-Booster}}

\newcommand{\myding}[1]{\ding{\numexpr181+#1\relax}}

\newcommand{\erase}[1]{}
\newcommand{\camera}[1]{}
\newcommand{\revision}[1]{{#1}}
\newcommand{\zypp}[1]{{}}

\newcommand\thintilde{{\lower.92ex\hbox{\mathtt{\char`~}}}}
\newcommand{\thicktilde}{{\lower.33ex\hbox{\texttt{\char`~}}}}
\definecolor{myblue}{HTML}{177cb0}      
\AtBeginDocument{%
  }

\copyrightyear{2025} 
\acmYear{2025} 
\setcopyright{acmlicensed}\acmConference[ISCA '25]{Proceedings of the 52nd
 Annual International Symposium on Computer Architecture}{June 21--25,
 2025}{Tokyo, Japan}
 \acmBooktitle{Proceedings of the 52nd Annual International Symposium on
 Computer Architecture (ISCA '25), June 21--25, 2025, Tokyo, Japan}
 \acmDOI{10.1145/3695053.3730987}
 \acmISBN{979-8-4007-1261-6/2025/06}

\begin{document}



\title{AIM: Software and Hardware Co-design for \underline{A}rchitecture-level \underline{I}R-drop \underline{M}itigation in High-performance PIM}


\author{Yuanpeng Zhang}
\affiliation{%
    \institution{School of Integrated Circuits}
    \institution{Peking University}
  \city{Beijing}
  \country{China}}
\affiliation{%
    \institution{Beijing Advanced Innovation Center for Integrated Circuits}
  \city{Beijing}
  \country{China}}
\email{zyp_cs@pku.edu.cn}
\orcid{0009-0007-4253-4747}

\author{Xing Hu}
\affiliation{%
    \institution{Houmo AI}
  \city{Nanjing}
  \country{China}}
\email{xing.hu@houmo.ai}
\orcid{0009-0003-4510-898X}

\author{Xi Chen}
\affiliation{%
    \institution{Southeast University}
  \city{Nanjing}
  \country{China}}
\email{xichen@seu.edu.cn}
\orcid{0009-0009-0119-5966}

\author{Zhihang Yuan}
\affiliation{%
    \institution{Houmo AI}
  \city{Nanjing}
  \country{China}}
\email{hahnyuan@gmail.com}
\orcid{0000-0001-7846-0240}

\author{Cong Li}
\affiliation{%
    \institution{School of Integrated Circuits}
    \institution{Peking University}
  \city{Beijing}
  \country{China}}
\affiliation{%
    \institution{Beijing Advanced Innovation Center for Integrated Circuits}
  \city{Beijing}
  \country{China}}
\email{leesou@pku.edu.cn}
\orcid{0000-0001-7760-3254}

\author{Jingchen Zhu}
\affiliation{%
    \institution{School of Integrated Circuits}
    \institution{Peking University}
  \city{Beijing}
  \country{China}}
\affiliation{%
    \institution{School of Computer Science}
    \institution{Peking University}
  \city{Beijing}
  \country{China}}
\email{zjc990112@pku.edu.cn}
\orcid{0000-0002-4321-7694}

\author{Zhao Wang}
\affiliation{%
    \institution{School of Integrated Circuits}
    \institution{Peking University}
  \city{Beijing}
  \country{China}}
\affiliation{%
    \institution{School of Computer Science}
    \institution{Peking University}
  \city{Beijing}
  \country{China}}
\email{wangzhao21@pku.edu.cn}
\orcid{0009-0009-8603-3850}

\author{Chenguang Zhang}
\affiliation{%
    \institution{School of Integrated Circuits}
    \institution{Peking University}
  \city{Beijing}
  \country{China}}
\affiliation{%
    \institution{School of Computer Science}
    \institution{Peking University}
  \city{Beijing}
  \country{China}}
\email{zhangchg@stu.pku.edu.cn}
\orcid{0000-0002-6476-7089}

\author{Xin Si}
\affiliation{%
    \institution{Southeast University}
  \city{Nanjing}
  \country{China}}
\email{xinsi@seu.edu.cn}
\orcid{0000-0002-4993-0087}

\author{Wei Gao}
\affiliation{%
    \institution{Houmo AI}
  \city{Shanghai}
  \country{China}}
\email{wei.gao@houmo.ai}
\orcid{0009-0006-6881-5224}

\author{Qiang Wu}
\affiliation{%
    \institution{Houmo AI}
  \city{Nanjing}
  \country{China}}
\email{qiang.wu@houmo.ai}
\orcid{0009-0009-8981-2876}

\author{Runsheng Wang}
\affiliation{%
    \institution{School of Integrated Circuits}
    \institution{Peking University}
  \city{Beijing}
  \country{China}}
\affiliation{%
    \institution{Beijing Advanced Innovation Center for Integrated Circuits}
  \city{Beijing}
  \country{China}}
\email{wrs@pku.edu.cn}
\orcid{0000-0002-7514-0767}

\author{Guangyu Sun}
\authornote{Corresponding author}

\affiliation{%
    \institution{School of Integrated Circuits}
    \institution{Peking University}
  \city{Beijing}
  \country{China}}
\affiliation{%
    \institution{Beijing Advanced Innovation Center for Integrated Circuits}
  \city{Beijing}
  \country{China}}
\email{gsun@pku.edu.cn}
\orcid{0000-0002-7315-6589}
\renewcommand{\shortauthors}{Yuanpeng Zhang et al.}


\begin{abstract} 
SRAM Processing-in-Memory (PIM) has emerged as the most promising implementation for high-performance PIM, delivering superior computing density, energy efficiency, and computational precision. However, the pursuit of higher performance necessitates more complex circuit designs and increased operating frequencies, which exacerbate IR-drop issues. Severe IR-drop can significantly degrade chip performance and even threaten reliability. Conventional circuit-level IR-drop mitigation methods, such as back-end optimizations, are resource-intensive and often compromise power, performance, and area (PPA). To address these challenges, we propose \name, comprehensive software and hardware co-design for architecture-level IR-drop mitigation in high-performance PIM. Initially, leveraging the bit-serial and in-situ dataflow processing properties of PIM, we introduce \tog\ and \HR, which establish a direct correlation between PIM workloads and IR-drop. Building on this foundation, we propose \lhr\ and \wds, enabling extensive exploration of architecture-level IR-drop mitigation while maintaining computational accuracy through software optimization. Subsequently, we develop \booster, a dynamic adjustment mechanism that integrates software-level HR information with hardware-based IR-drop monitoring to adapt the V-f pairs of the PIM macro, achieving enhanced energy efficiency and performance. Finally, we propose the HR-aware task mapping method, bridging software and hardware designs to achieve optimal improvement. Post-layout simulation results on a 7nm 256-TOPS PIM chip demonstrate that \name\ achieves up to 69.2\% IR-drop mitigation, resulting in 2.29$\times$ energy efficiency improvement and 1.152$\times$ speedup.
\end{abstract}

\ccsdesc[500]{Computer systems organization~Architectures}
\ccsdesc[500]{Hardware~Power and energy}
\ccsdesc[300]{Computing methodologies~Machine learning}
\ccsdesc[100]{Security and privacy~Security in hardware}


\keywords{Processing-in-memory, IR-drop mitigation, Architecture-level optimization, Software and hardware co-design}


\maketitle

\section{Introduction} \label{sec:Introduction}
Processing-in-memory (PIM) circuit seamlessly combines storage and computing functions, embedding computational capabilities within the memory to enhance data processing efficiency~\cite{CIM1,CIM0,CIM2,CIM3,CIM4}. Compared to eNVMs and DRAM, SRAM-based PIM (SRAM PIM) offers higher computational precision and achieves superior energy efficiency and computing density by leveraging state-of-the-art (SOTA) CMOS process technologies~\cite{SRAM-CIM-survey0,SRAM-CIM-survey1,DCIM0,DCIM-TSMC,DCIM-MediaTek,SRAM-PIM-survey0}. For instance, TSMC's latest SRAM PIM design, utilizing 3nm process technology, achieves $55.0\ TOPS/mm^2$ and $32.5\ TOPS/W$ while supporting INT12 $\times$ INT12 computations~\cite{DCIM-TSMC}.  Consequently, SRAM PIM has been extensively studied in academia as a computing paradigm with substantial development potential~\cite{DCIM0, DCIM-TSMC, DCIM3, DCIM4, DCIM5, DCIM6, DCIM7, DCIM8, DCIM10, DCIM9,SRAM-CIM1,SRAM-CIM} and is widely regarded as the leading candidate for commercial high-performance PIM chips. Numerous companies and startups have developed commercial SRAM PIM products for various AI scenarios. Examples include d-matrix~\cite{dmatrix} for large language model serving, Houmo~\cite{houmo} for autonomous driving, MediaTek~\cite{DCIM-MediaTek} and PIMCHIP~\cite{pimchip} for mobile devices, as well as Axelera~\cite{Axelera}, Synthara~\cite{Synthara}, and Surecore~\cite{Surecore} for IoT applications, etc. Among these, d-Matrix's Corsair platform is specifically designed for generative AI applications in data centers, delivering up to 9600-TFLOPS computational power while offering superior power efficiency, reduced latency, and lower total cost of ownership (TCO)~\cite{dmatrix}. Similarly, Houmo's MoMagic30~\cite{Momagic}, an edge AI chip, delivers up to 100-TOPS computing power with a typical power consumption of 12W, supporting mainstream large language models (LLMs) and achieving performance of 15\thicktilde20 tokens per second. Additionally, MediaTek has integrated TSMC's 3nm SRAM PIM Engine into its smart devices~\cite{DCIM-MediaTek}. These advancements underscore the promise of SRAM PIM as a high-performance solution for efficient AI processing.

Nevertheless, during the design and signoff process of a high-performance PIM chip, we found that it faces the challenge of severe {\bf IR-drop}. IR-drop is a phenomenon where the supply voltage is influenced by current within a functioning circuit, resulting in increased clock delay and even reliability failures~\cite{IRdrop1,IRdrop2,IRdrop3}. This problem is a widespread issue in all types of chips, including AI processors, CPUs, and GPUs~\cite{circuitnet2,circuitnet1,DVFS0-CIM,IR-CPU0,IR-CPU1}. And it is especially pronounced in high-performance PIM chips, where the simultaneous operation of numerous computing logic units within the same cycle generates substantial instantaneous current. Typically, hardware engineers rely on various circuit-level IR-drop mitigation strategies to ensure functionality and safety. These strategies include cell relocations, modifications to power planes, and the addition of decoupling capacitors~\cite{ECO0,ECO1,ECO2,ECO3}. However, such methods require significant time and resources, often compromising the chip’s performance, power, and area (PPA)~\cite{ECO-cost0,ECO-cost1,ECO-cost2}.  For example, the Graphcore Bow IPU~\cite{ipu} employs wafer-on-wafer (WoW) 3D packaging technology and incorporates deep trench capacitors (DTCs) in the power delivery die near the computing units to mitigate a 100mV IR-drop~\cite{wow}, resulting in great design cost and energy consumption. Similarly, we observed an even higher IR-drop of 140mV in our commercially available 7nm 256-TOPS SRAM PIM chip. These challenges highlight the critical need for architecture-level IR-drop mitigation methods to reduce design costs while improving chip performance and power efficiency.

Architecture-level IR-drop mitigation requires establishing a clear relationship between IR-drop and workload characteristics. Unlike general-purpose computing units, PIM architectures possess unique opportunities in this regard. First, PIM workloads are inherently more regular and often pre-determined. PIM chips are primarily designed for neural network workloads and, in specific scenarios such as autonomous driving~\cite{autonomousdriving0,autonomousdriving1}, even only execute a limited set of specialized models. It simplifies the task of analyzing workload characteristics. Second, PIM naturally supports in-situ processing~\cite{SRAM-CIM-survey0,SRAM-PIM-survey0}, enabling the decoupling of input and weight data, thereby allowing optimizations without being affected by user input variability. Leveraging these characteristics, we proposed AIM, a comprehensive software and hardware co-design for \underline{A}rchitecture-level \underline{I}R-drop \underline{M}itigation on high-performance PIM. To the best of our knowledge, we are the first to mitigate IR-drop in high-performance PIM from an architectural perspective. Our contributions are multifaceted:
\begin{itemize}[leftmargin=16pt]
\item We propose \tog\ \revision{(instantaneous toggle rate of bitstreams in a PIM bank)} and \HR\ \revision{(average hamming rate of weights)}, two architecture-level metrics correlating workloads and IR-drop.
\item We introduce \lhr\ (lower hamming rate), a regularization term to reduce \HR\ by penalizing high-HR weights with negligible accuracy loss during quantization. The code\footnote{\href{https://github.com/pku-zyp/LHR-of-AIM-in-ISCA25.git}{https://github.com/pku-zyp/LHR-of-AIM-in-ISCA25.git}} for integrating \lhr\ with quantization algorithms has been open-sourced.
\item We present \wds\ (weight distribution shift), a software method leveraging data distribution to further reduce \HR.
\item We develop \booster, a dynamic V-f pairs adjustment method combining software-derived HR information and hardware-based IR-drop monitoring to achieve IR-drop mitigation while improving energy efficiency and performance.
\item We introduce \HR-aware task mapping, a strategy that incorporates \HR\ when mapping tasks to hardware units controlled by \booster, achieving optimal performance of software and hardware collaboration.
\end{itemize}

We validate \name\ using post-layout simulation data obtained from RedHawk~\cite{TOOL_redhawk-sem} and HSPICE~\cite{TOOL_hspice} on a 7nm 256-TOPS PIM chip \revision{design}. Experimental results demonstrate that \name\ achieves up to 69.2\% IR-drop mitigation, resulting in up to 2.29$\times$ energy efficiency improvement or 1.152$\times$ performance improvement.

\section{Background}  \label{sec:background}
\subsection{SRAM PIM} \label{subsec:PIM}
\begin{figure}[t]
    \centering
    \includegraphics[width=\linewidth]{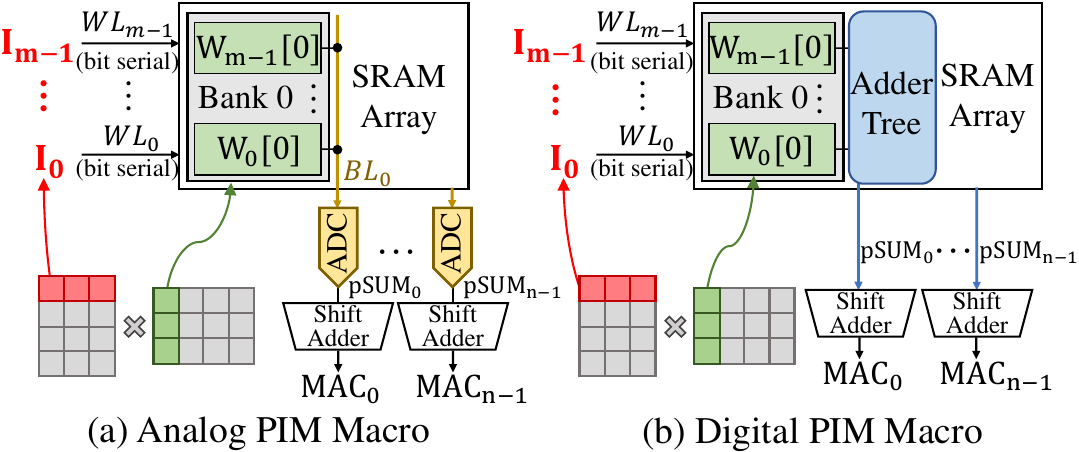}
    \caption{SRAM PIM and data placement} 
    \label{fig:SRAM PIM}
\end{figure}

SRAM PIM has emerged as a promising architecture for high-performance computing, offering exceptional computing density, energy efficiency, and computational precision. These advantages are rooted in SRAM's rapid access times (\textasciitilde 1ns read/write latency), high endurance, and utilization of state-of-the-art CMOS process technologies~\cite{SRAM-CIM-survey0,SRAM-CIM-survey1,SRAM-CIM1,SRAM-CIM4}. In SRAM PIM architectures, many multiplicands stored in SRAM cells engage in bit-wise multiplication with input multipliers, with subsequent accumulation performed through dedicated paths. Throughout the entire matrix-matrix multiplication process, the data in SRAM cells remains unchanged, and input data is loaded bit-serially, referred to as \textbf{in-situ processing}. This approach substantially reduces energy consumption while achieving fast multiply-accumulate (MAC) operations. SRAM PIM architectures can be categorized into two types based on their data accumulation format: Analog PIM (APIM)~\cite{SRAM-ACIM0} and Digital PIM (DPIM)~\cite{DCIM-MediaTek}. As shown in Figure~\ref{fig:SRAM PIM}-(a), in APIM, the products are accumulated as an analog bit-line (BL) voltage, which is subsequently converted to a digital value by an analog-to-digital converter (ADC). In contrast, DPIM relies entirely on digital representation, where data processing and computation occur via the outputs of logic gates and adder trees (Figure~\ref{fig:SRAM PIM}-(b)). This digital approach minimizes the effects of Process-Voltage-Temperature (PVT) variations, thereby improving operational reliability~\cite{DCIM0,DCIM-MediaTek,DCIM3,DCIM4,DCIM5,DCIMvsACIM}. 

Although APIM and DPIM differ in the form of product accumulation, they share similarities in data placement and processing paradigm. As shown in Figure~\ref{fig:SRAM PIM}, this paper designates the multiplied data stored in SRAM as \textbf{in-memory data} (denoted as W) and the serialized data sequentially loaded on the word line (WL) as \textbf{input data} (denoted as I). Specifically, for operations involving fixed weights, such as convolution layers (conv), weights are typically in-memory data, while features are loaded as input data. Conversely, for variable-operand operators such as QK\textsuperscript{T} or SV in transformer attention, either matrix can act as in-memory data, depending on the specific configuration and shape of the operation.

\subsection{A Brief Introduction to IR-drop}\label{sec:IRdrop}

\begin{figure}[t]
    \centering
    \includegraphics[width=0.92\linewidth]{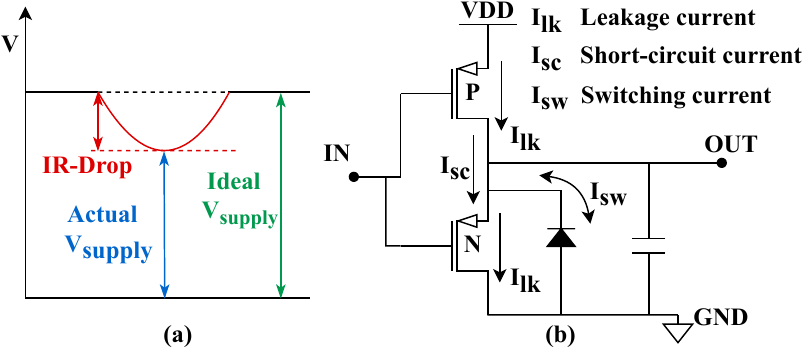}
    \caption{(a) IR-drop (b) Static and dynamic current} 
    \label{fig:IR-drop}
\end{figure}

IR-drop represents the voltage discrepancy between the ideal supply voltage and the actual voltage received by each circuit cell (Figure~\ref{fig:IR-drop}-(a)).  The magnitude of the IR-drop is positively correlated with the product of current (I) passing through the power delivery network (PDN) parasitic resistance (R). IR-drop can be classified into static IR-drop and dynamic IR-drop, distinguished by the origins and behaviors of corresponding currents.

Static IR-drop is caused by static current. Static current is drawn by an integrated circuit when it is idle, unloaded, and not actively switching, but still enabled. It is generally equivalent to the leakage current, denoted as $I_{lk}$ in Figure~\ref{fig:IR-drop}-(b), which represents the gradual flow of electrical charge across boundaries that are typically viewed as insulating, such as current leakage through transistors in the ``off'' state. Static IR-drop is primarily influenced by the PDN design, as well as the placement and routing of the circuit.

Dynamic IR-drop is caused by dynamic current, which occurs when the integrated circuit is in operation. It mainly consists of two types of current: short-circuit current and switching current. As illustrated in Figure~\revision{\ref{fig:IR-drop}-(b)}, short-circuit current, denoted as $I_{sc}$, arises when there is an instantaneous short-circuit connection between the supply voltage and ground during a gate state transition. Switching current, represented as $I_{sw}$, is produced through the charging and discharging of internal and net capacitances. Dynamic IR-drop is thus highly sensitive to runtime activities, including gate switching and the capacitive charge and discharge cycles.
\section{Motivation} \label{sec:Motivation}

\subsection{IR-drop Impact and Circuit-level Mitigation}

As chip performance advances and frequency increases, the IR-drop problem is becoming more and more serious. For digital chips, including DPIM, CPUs, GPUs, and TPUs~\cite{circuitnet2,circuitnet1,DVFS0-CIM,IR-CPU0,IR-CPU1}, significant IR-drop reduces the voltage to standard cells, impairing performance and leading to time violations such as setup$/$hold violations~\cite{IRdrop1,IRdrop2,IRdrop3}. Furthermore, a severe IR-drop that prevents a cell from meeting the minimum operational voltage can result in the chip's functional failure~\cite{IRdrop4}. For analog chips, like APIM, IR-drop directly affects the BL voltage used for calculations, degrading energy efficiency and computational accuracy~\cite{APIM-IR0,APIM-IR1,APIM-IR2}. 

To mitigate these effects, IC designers are compelled to adopt numerous time-intensive and costly circuit-level mitigation techniques, often compromising the chip’s performance, power, and area (PPA) metrics~\cite{ECO0,ECO1,ECO2,ECO3}. These interventions include widening power supply lines, increasing the density of power switches, adjusting switch placement within floorplan channels and boundaries~\cite{IRsolve1,IRsolve3,ECO-cost1,IRsolve5,IRsolve6}, inserting decoupling capacitors~\cite{IRsolve0}, and managing timing slack by optimizing clock tree structures~\cite{IRsolve2}. In spite of these efforts, careful clock frequency planning remains necessary to prevent timing violations induced by IR-drop under extreme conditions. Moreover, we found that although the circuit-level mitigation techniques have a good effect on static IR-drop, it is too pessimistic when mitigating dynamic IR-drop.

\subsection{Opportunity and Challenges}
\begin{figure}[t]
    \centering
    \includegraphics[width=0.95\linewidth]{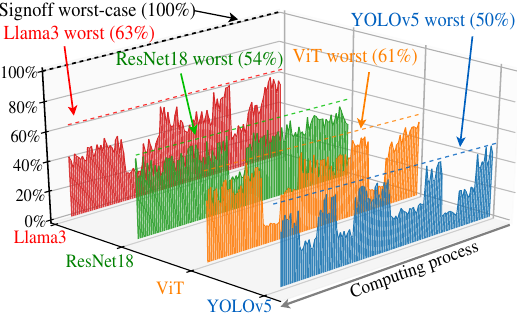}
    \caption{Normalized IR-drop at different workloads} 
    \label{fig:IR-workload}
\end{figure}

To ensure chip reliability and prevent IR-drop violations under any workload, IC designers typically employ rigorous testbenches for chip design and signoff~\cite{worse-IR0,worse-IR1,worse-IR2,worse-IR3}. This includes evaluating the IR-drop under the theoretically most demanding workload, recorded as the signoff worst-case IR-drop.
However, circuit-level IR-drop mitigation techniques targeting the worst-case IR-drop are too pessimistic, revealing the opportunity for optimization for two main reasons.
First, unlike general-purpose computing cores, high-performance PIM architectures are specifically tailored to support AI applications, making their workloads more predictable and deterministic. By sequentially mapping operators from several workloads onto a 256-TOPS SRAM PIM and profiling the IR-drop during processing, we observe significant discrepancies compared to the signoff worst-case. As illustrated in Figure~\ref{fig:IR-workload}, while IR-drop fluctuates significantly during processing, the worst IR-drop of different workloads remains considerably lower than the signoff worst-case. Moreover, we find that the worst IR-drop for each model exhibits a relatively stable pattern with different input images or texts. These phenomena indicate a potential correlation between workloads and IR-drop.
Second, PIM naturally supports in-situ dataflow processing and bit-serial multiplication, resulting in an observable data behavior model. By identifying and exploiting the relationship between workloads and IR-drop, architecture-level IR-drop mitigation strategies can be developed to avoid the pessimism in circuit-level IR-drop mitigation.

Despite the opportunity, there are still several challenges to be solved. 1) The foundation of architecture-level IR-drop mitigation lies in identifying the correlation between workload and IR-drop and expressing it in a statistically manageable form. Blindly testing all workloads and measuring IR-drop is neither cost-effective nor feasible, given the vast number of models and the large size of datasets. 2) Profiling results show that the model itself accounts for only about 30\%-50\% of the available architectural IR-drop mitigation space, necessitating low-cost software methods to further expand this space. 3) The available mitigation space varies across models, and even the IR-drop can differ significantly between operators within the same model. This requires software-guided hardware dynamically supporting various IR-drop degrees. 
\section{Correlating Workload and IR-drop} \label{sec:prep}
In this section, we lay the groundwork for \name. Building on the definition and characteristics of IR-drop, we propose \tog, \revision{which represents the instantaneous toggle rate of bitstreams from SRAM cells to the adder component in a PIM bank.} This metric serves as an architecture-level indicator that quantitatively captures the correlation between IR-drop and workload. We validate the effectiveness of \tog\ through experiments. Additionally, we propose HR, \revision{defined as the average hamming rate of quantized weights}, as a simplified and optimizable metric derived from \tog, which is input data independent. HR enables more effective exploration of architecture-level IR-drop mitigation strategies.

\subsection{\tog: an Architecture-level IR-drop Indicator}
Many previous works~\cite{IRdrop0,IRdrop1,IRdrop2,IRdrop5} and commercial tools~\cite{TOOL_redhawk-sem,TOOL_hspice}  commonly use gate-level bit toggle rates to evaluate circuit IR-drop. These tools model the power network and combine it with toggle rates obtained from logic simulations to simulate current distributions and calculate the IR-drop across components.

Inspired by these circuit-level IR-drop estimation tools, which correlate dynamic current through components with nearby circuit activities, we propose \tog, an architecture-level indicator linking workload characteristics directly to IR-drop. \tog\ quantifies the cycle-to-cycle toggle rate of bitstreams from SRAM cells to the adder component in a PIM bank. Given in-memory data $\{W_k\}_{k=1}^n$ stored in a PIM bank with $n$ cells. Each $W_k$ is quantized to $q$ bits and the bit values are denoted as $W_{k,1}, \dots, W_{k,q}$. The input data at cycle $t$ and $t+1$ are $\{I_{k,t}\}_{k=1}^n$ and $\{I_{k,t+1}\}_{k=1}^n$. \tog\ of a PIM bank in cycle $t$ is expressed as the Equation~\ref{equ:R_ins}:
\begin{equation}
    \text{in cycle}\ t,\ \tog = \frac{\sum_{k=1}^n\sum_{i=1}^{q}(W_{k,i}\land(I_{k,t}\oplus I_{k,t+1}))}{nq}
    \label{equ:R_ins}
\end{equation}
The IR-drop can be roughly estimated by Equation~\ref{equ:IRdrop}:
\begin{equation}
    \begin{split}
	IR\text{-}drop &= \Delta V_{static} + \Delta V_{dynamic}\\
	\Delta V_{static} &\approx k_{lk}I_{lk}R_{lk}\\
	\Delta V_{dynamic} &\approx (k_{sc}I_{sc}R_{sc}+k_{sw}I_{sw}R_{sw})\times\textbf{\tog}
    \end{split}
    \label{equ:IRdrop}
\end{equation}
where $I_{lk}$, $I_{sc}$, and $I_{sw}$ represent leakage, short-circuit, and switching currents, respectively, while $R_{lk}$, $R_{sc}$, and $R_{sw}$ are the corresponding resistances. $k_{lk}$, $k_{sc}$, and $k_{sw}$ are constant coefficients.

\begin{figure}[t]
    \centering
    \includegraphics[width=0.98\linewidth]{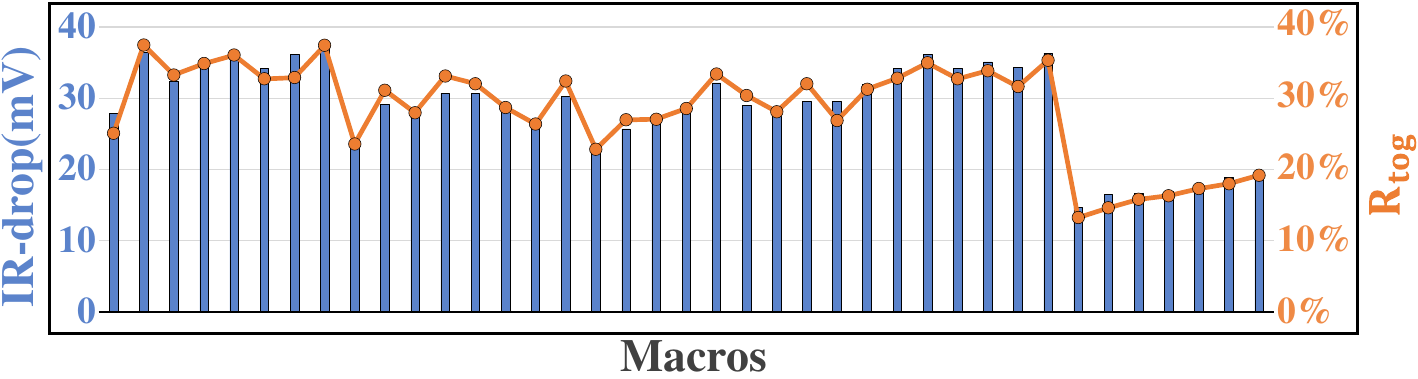}
    \caption{Correlation of IR-drop and \tog} 
    \label{fig:Correlation}
\end{figure}
Unlike circuit-level tools that require complex modeling and simulations, \tog\ simplifies IR-drop estimation by treating the PIM bank as a whole area with relatively stable equivalent resistance. While this approach sacrifices precise per-component estimations, \tog\ serves as a robust indicator for maintaining a partial order relationship with IR-drop, aligning well with intuition: higher \tog\ typically reflects increased IR-drop activities.

To validate \tog, we conducted experiments on 7nm DPIM macro and 28nm APIM macros, estimating IR-drop using HSPICE~\cite{TOOL_hspice}. The average \tog\ across all banks in a PIM macro correlates strongly with IR-drop, achieving coefficients of 0.977 for DPIM and 0.998 for APIM across various workloads. Figure~\ref{fig:Correlation} shows the linear correlation results between IR-drop and \tog\ of the DPIM macros. \camera{Therefore, unlike gate-level toggle rates in EDA tools which are unreadable and hard to get, \tog\ offers straightforward assessment opportunities as a workload-dependent indicator.}
 
\subsection{\HR: an Optimizable Metric for Mitigation} \label{subsec:HR}
Although \tog\ is an accurate architecture-level indicator, it varies dynamically with workloads, especially input streams. \tog\ can exceed $10^8$ toggles for a single input image. It makes real-time software analysis infeasible due to resource and practicality constraints. 

\begin{figure}[t]
    \centering
    \includegraphics[width=0.98\linewidth]{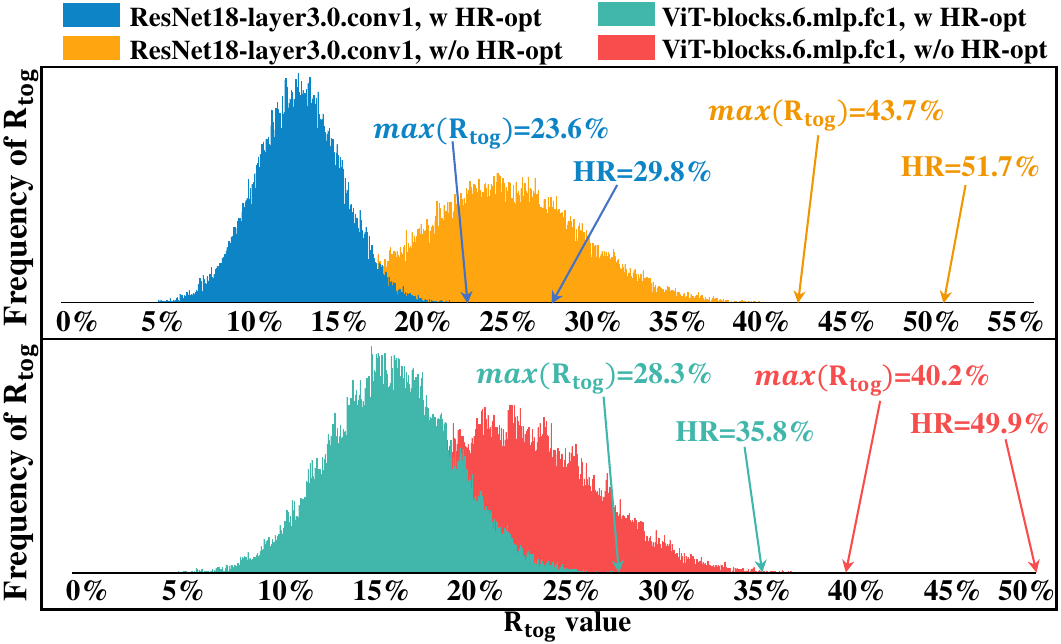}
    \caption{\revision{\tog\ distribution: HR dominates the maximum of \tog\ and HR optimization reduces overall \tog}} 
    \label{fig:tog distribution}
\end{figure}

To address this, we introduce \HR\ (Hamming Rate), an optimizable metric derived from \tog. The core idea behind \HR\ is to leverage the in-situ processing and bit-serial properties of PIM architectures, decoupling the pre-loaded data (network weights) from the dynamic input feature streams. \camera{This reduces both the average and peak \tog\ values, thus mitigating IR-drop. }HR is defined as Equation~\ref{equ:hamming}.

\begin{equation}
    \begin{split}
        &HM(\{W_n\}) := {\sum_{i=1}^{q} \sum_{k=1}^n W_{k,i}}\\
        &HR(\{W_n\}):=\frac{HM(\{W_n\})}{nq}
    \end{split}
    \label{equ:hamming}
\end{equation}

Where $HM(\{W_n\})$ represents the hamming value of $\{W_n\}$, counting all valid bits (1s) in $\{W_k\}$. \HR\ is the hamming value divided by $nq$, determined only by in-memory data (W). 

\begin{gather}
    \begin{split}
    &\because\ \forall k,\ \forall t, s_{k,t}\oplus s_{k,t+1}=\{0, 1\}\leqslant 1\\
    &\therefore \tog = \frac{\sum_{k=1}^n\sum_{i=1}^{q}(W_{k,i}\&(I_{k,t}\oplus I_{k,t+1}))}{nq}\leqslant\frac{\sum_{k=1}^n\sum_{i=1}^{q}W_{k,i}}{nq}\\
    &\therefore\  sup(\tog)= \frac{\sum_{k=1}^n\sum_{i=1}^{q}W_{k,i}}{nq}=HR(\{W_n\})
    \end{split}
    \label{equ:R_sup}
\end{gather}

Equation~\ref{equ:R_sup} establishes that \camera{\tog\ at any given time is governed by two factors: the number of weights set to 1 and the simultaneous bit-flips in input streams. Consequently, }\HR\ represents the theoretical upper bound of \tog\ when all input bitstreams toggle concurrently. Experiment results further support this. As shown in Figure~\ref{fig:tog distribution}, we profiled the \tog\ distribution of a ResNet18/ViT operator over 50,000 cycles (consistent trends hold across other layers and networks). Notably, the observed peak \tog\ never exceeds the corresponding \HR\ value, with a substantial margin between the two. Moreover, optimizing \HR\ helps reduce the overall value of \tog.

Therefore, by optimizing the pre-loaded in-memory data, \HR\ allows us to mitigate IR-drop without accounting for input stream variations, offering a practical, architecture-level solution.

\section{\name\ Design} \label{sec:design}
\subsection{\name\ Overview} \label{sec:overview}
\begin{figure}[t]
    \centering
    \includegraphics[width=0.94\linewidth]{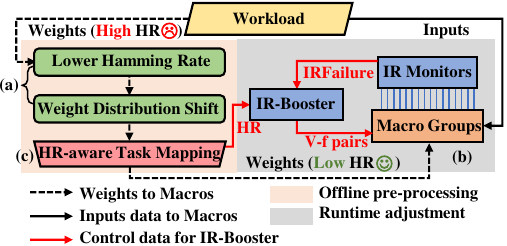}
    \caption{\name\ design and workflow: \revision{(a) Offline software-based HR optimization methods: \lhr\ and \wds\ (b) A runtime adjustment hardware method: IR-Booster (c) An interconnection and co-optimization method: HR-aware task mapping}} 
    \label{fig:workflow}
\end{figure}
As illustrated in Figure~\ref{fig:workflow}, \name\ is structured into two complementary components: offline software-based HR optimization and a runtime hardware adjustment method guided by software. These components are interconnected through HR-aware task mapping to enable cohesive system-level optimization. In Section~\ref{subsec:integration}, we first outlined the basic implementation and usage of each component, and provided an end-to-end example to demonstrate their synergy. We then detailed the design principles and methodologies of each component. Specifically, in Section~\ref{sec:lhr}, we introduce \lhr, a lightweight regularization term integrated into the quantization process. This term reduces the HR of weights during pre-processing, improving robustness without significantly impacting accuracy. In Section~\ref{sec:wds}, building on post-quantization data distribution and its HR characteristics, we propose \wds, a method that refines the overall distribution to further reduce the weighted HR of the network.
In Section~\ref{sec:booster}, we present \booster, a runtime hardware-level adjustment mechanism that combines software-derived HR information with hardware-based IR-drop monitoring. By dynamically selecting optimal voltage-frequency pairs for each PIM macro, \booster\ mitigates IR-drop at the architectural level while enhancing energy efficiency and performance. Finally, in Section~\ref{sec:Mapping}, we propose HR-aware task mapping, which assigns segmented tasks to PIM macros managed by \booster. It maximizes \booster's effectiveness, further optimizing overall system performance.
\subsection{\revision{Application Integration and Usage}} \label{subsec:integration}

\subsubsection{\revision{\name\ component: implementation and selection}}\label{subsubsec:independent}\ \par

\revision{Each method of \name\ is implemented independently, offering flexibility for selective use based on user needs.}
\begin{enumerate}[labelsep = .5em, leftmargin = 0pt, itemindent = 3em]
\item \revision{\lhr: It integrates easily into the application layer by adding a term to training loop (e.g., loss += $\lambda *$lhr\_norm(model.parameters()), in PyTorch). It requires no modifications to underlying frameworks (such as PyTorch, TensorFlow, or JAX) due to its gradient-based design, requiring only standard API calls without low-level hacking.}

\item \revision{\wds: Integrated within the compiler, \wds\ has a default $\delta$ value of 8. However, users can explicitly specify different $\delta$ values for each operator supported by the compiler.}
\item \revision{\booster: Directly integrated into the PIM chip as a hardware technique, \booster\ operates independently of \lhr\ when fine-tuning is not feasible (e.g., due to privacy concerns). \booster\ alone can provide significant energy efficiency improvements, with \lhr\ and \wds\ providing even greater benefits.}
\item \revision{HR-aware Task Mapping: Integrated within the compiler, it is applied after determining the operator scheduling and segmentation. Its primary goal is to reduce performance degradation caused by mutual interference when networks with different HR values are mapped onto the PIM chip simultaneously. Since it involves multiple rounds of simulation introducing second-level latency, this method should be used cautiously. For complex workloads (especially those with various operators or networks), we recommend utilizing HR-aware Task Mapping for optimal IR-drop mitigation.}
\end{enumerate}

\subsubsection{\revision{Synergy of all \name\ components: an end-to-end example}}\label{subsubsec:synergy}\ \par
\revision{All the software methods and hardware technologies in \name\ work in concert to deliver optimal improvements. We present an end-to-end example of how they collaborate.}

\revision{During the processing phase, \name\ begins by loading the target network weights, which have been optimized using the quantized architecture enhanced with \lhr. The compiler then reads the $\delta$ configuration for each operator and performs \wds\ operations to adjust weights accordingly. Next, the compiler applies a simulated annealing-based task mapping strategy, using the HR values obtained during processing and input data randomly drawn from standard datasets (e.g., ImageNet~\cite{imagenet}, Wikitext2~\cite{wikitext}). These HR values corresponding to all mapped tasks are passed to \booster.}

\revision{Once the compilation is complete and the optimized weights are loaded, \name\ enters the inference phase. During runtime, \booster\ dynamically selects V-f pairs based on the user's selected operating mode. It also continuously monitors for IRFailures and triggers recalculation when necessary. This integrated approach allows \name\ to achieve substantial energy efficiency and processing performance improvements while maintaining reliability.}

\subsection{Lower Hamming Rate} \label{sec:lhr}
In this section, we introduce \lhr\ (lower hamming rate), a regularization to decrease the \HR\ throughout the entire network with minimum accuracy influence during quantization. By targeting the layers with the highest \HR, \lhr\ aims to mitigate IR-drop at the network level. Unlike many energy-aware quantization methods, \lhr\ prioritizes reducing the \HR\ of critical layers over only achieving a theoretically minimal overall average \HR.

Regularization terms are commonly added to loss functions during Quantization Aware Training (QAT) to enforce additional constraints. A straightforward approach to reducing \HR\ would involve applying the $L_1$ norm to penalize weights with higher \HR. However, \HR\ is a non-differentiable integer metric, making it unsuitable for backpropagation. To overcome this, we approximate the \HR\ of a floating-point weight $w$ using a linear interpolation between its two nearest integer values. This allows us to define a differentiable approximation of \HR, enabling its use in backpropagation.

\begin{equation}
    \begin{split}
        &low = \lfloor \cfrac{w}{s_w} \rfloor,\ high = \lceil \cfrac{w}{s_w} \rceil,\ p = \cfrac{w}{s_w} - low\\
        &HR(w) = (1-p)\cdot HR[low] + p\cdot HR[high]
    \end{split}
    \label{equ:Interpolation}
\end{equation}

Given a floating-point weight $w$ and its quantization scale $s_w$, we determine the nearest integers $low$ and $high$. Using $HR[\ ]$ to represent the HR value of the corresponding integer, we calculate interpolated \HR\ of floating-point weight $w$ as Equation~\ref{equ:Interpolation}.

\begin{figure}[t]
    \centering
    \includegraphics[width=\linewidth]{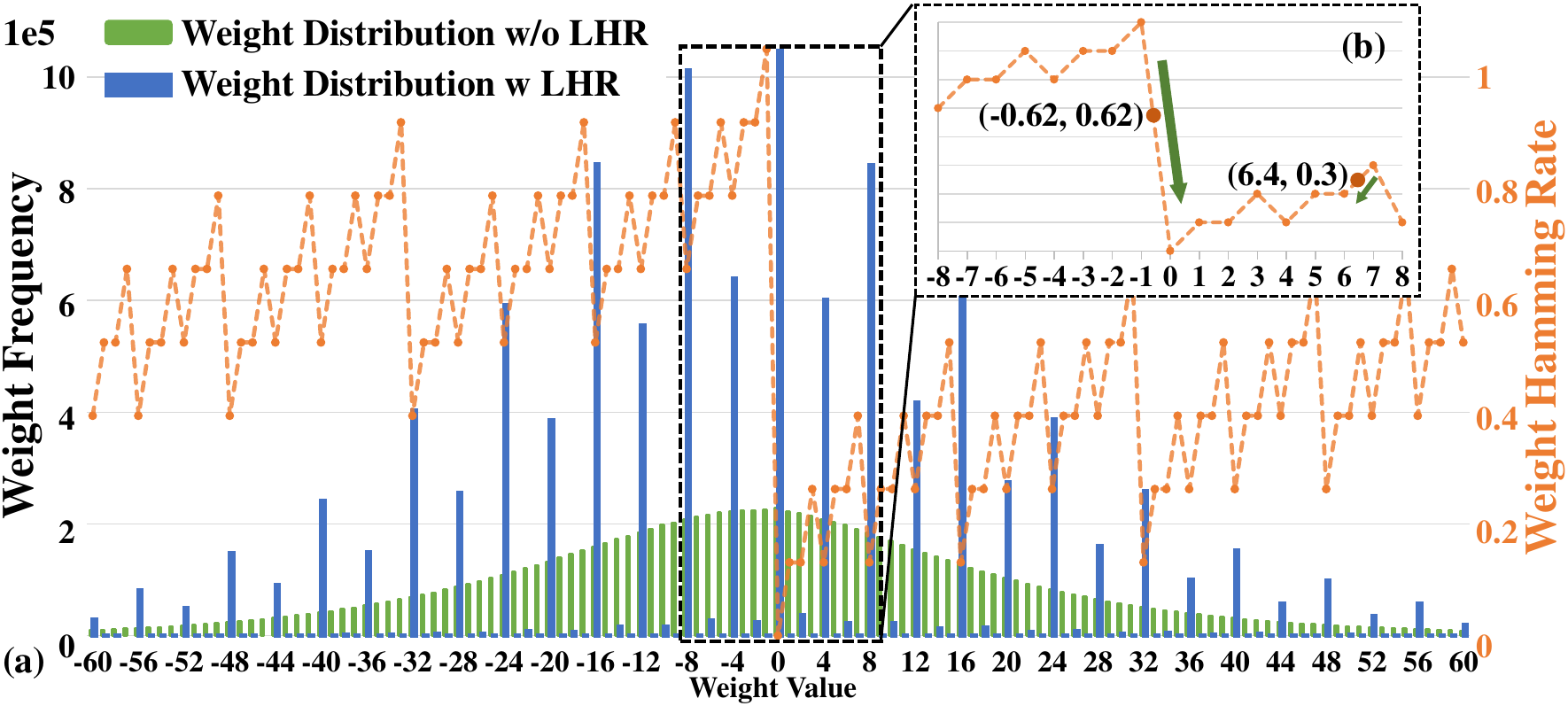}
    \caption{\revision{(a) Weight distribution with \lhr\ aligns with local minima
of the Hamming Rate (b) Calculating the HR and corresponding gradient of a floating-point by interpolation}}
    \label{fig:Distribution}
\end{figure}

During backpropagation, the gradient of $w$ is derived from the slope of the interpolated line segment. This gradient encourages weights to converge toward regions with lower \HR.
For example, in Figure~\ref{fig:Distribution}-(b), the interpolated \HR\ of $-0.62$ is $0.62$, with a gradient of $1$, while the \HR\ of $6.4$ is $0.3$, with a gradient of $-0.125$. Larger differences in \HR\ between adjacent integers lead to steeper gradients, enabling faster convergence to optimal values.


\lhr\ can be incorporated into either quantization fine-tuning or directly into quantization algorithms. For a classification task, the total loss function with \lhr\ is expressed as:
\begin{equation}
    \begin{split}
        &L_{HR} = \sum_{i=1}^n HM_{rate}^2(layer_i)\\
        &L_{all}=L_{Task}+\lambda L_{HR}
    \end{split}
    \label{equ:loss}
\end{equation}

Where $L_{Task}$ represents the original task-specific loss (e.g., cross-entropy for classification), and $\lambda$ is a regularization parameter used to balance the \HR\ reduction and accuracy loss. $L_{HR}$ represents the hamming loss of the network, calculated by the squared sum of the average \HR\ for each layer in the entire network. This penalizes layers with higher \HR, thereby reducing both the overall network's \HR\ and especially the peak \HR\ across all layers.

To validate \lhr, we profile the quantized weight distribution of ResNet18. As shown in Figure~\ref{fig:Distribution}-(a), compared to the widely adopted quantization method~\cite{quant-baseline} as a baseline without \HR\ considering, the introduction of \lhr\ encourages weights to align with local minima of the Hamming function, such as -8, 0, and 8. This alignment reduces \HR\ of weights, effectively mitigating IR-drop.
\subsection{Weight Distribution Shift} \label{sec:wds}
While \lhr\ reduces the overall \HR\ by encouraging weights to converge at local minima of HR, quantization generally preserves the relative sign (positive or negative) of weights~\cite{round-adaround,round-GPTQ,round-intel,round-OmniQuant}. This means weights remain normally distributed, with more weights concentrated near zero and small magnitudes (e.g., -8 and 8), as shown in Figure~\ref{fig:Distribution}. According to the rules of two’s complement encoding, positive values with smaller absolute magnitudes have lower \HR, whereas negative values with smaller absolute magnitudes exhibit higher \HR. This observation motivates us to propose weight distribution shift (\wds), which reduces overall HR by shifting the data distribution towards positive values, concentrating weights around smaller positive numbers.

\subsubsection{\WDS\ Methodology}
\ \par

\begin{algorithm}[t]
    \caption{Weight distribution shift}
    \label{algo:shift}
    \begin{algorithmic}[1]
        \FOR{$Layer\ in\ \{Layers\}$} 
            \STATE \textcolor{OliveGreen}{\textit{/* Offline preprocessing (Outside Critical Path): */}}
            \FOR{$Weight\ in\ Layer$}
                \STATE $Weight' = IfOverflow(INTMAX_{nbit}, (Weight + \delta))$
            \ENDFOR
            \STATE \textcolor{red}{\textit{/* MM multiplication with lower HR (On Critical Path): */}}
            \STATE $Output = Matmul(Layer',\ Input)$
            \STATE \textcolor{MidnightBlue}{\textit{/* Shift compensation (Outside Critical Path): */}}
            \STATE $Correc = -Sum(Input)\times\delta$, $Output = Output + Correc$
        \ENDFOR
    \end{algorithmic}
\end{algorithm}
Shifting the weight distribution towards positive values can be implemented by adding a constant value, $\delta$, to all weights offline\camera{ before loading them into the PIM chip}. 
This adjustment occurs immediately after quantization. Although this shift reduces the \HR\ of weights used for matrix multiplications, it introduces numerical errors that can significantly degrade the accuracy of the neural network. Therefore, it is essential to compensate for these errors after the matrix multiplication operation.\camera{ , ensuring the correction is performed outside the critical path to avoid delaying computation.} Algorithm~\ref{algo:shift} illustrates the process of \wds. 

1) Offline preprocessing: To minimize IR-drop during matrix-matrix multiplication (line 7), $\delta$ is added to weights offline before they are loaded (lines 3-5). If the resulting weight value exceeds the maximum representable value for the quantization bitwidth (e.g., INT8), it is clamped to $INTMAX$. This avoids overflow into negative values, which could otherwise cause unknown errors and degrade inference accuracy. Profiling reveals such overflows occur in less than 1\% of weights, minimizing their impact. When selecting $\delta$ values, we notice that weights processed through quantization with \lhr\ often cluster at local minima with lower hamming, such as 0, -8, or 8 (Figure~\ref{fig:Distribution}-(a)). Therefore, $\delta$ must be chosen to align with this pattern; improper selection could inadvertently increase $HR$. For INT8 quantization, $\delta$ values of 8 or 16 produce better results. While for INT4 quantization, $\delta$ values of 2 or 4 are more suitable.

2) Shift compensation: Errors introduced by $\delta$ are corrected post-matrix multiplication. First, the $Correction$ value is calculated by $Input$ and $\delta$ (line 9). Since all weights in a layer share the same $\delta$, the $Input$ values are summed before being multiplied by $\delta$ to simplify hardware and reduce multiplications. Subsequently, in line 10, the $Correction$ is added to the $Output$ to rectify the calculation errors. Detailed hardware support for shift compensation is as follows.


\subsubsection{Shift Compensator Hardware Design} \label{subsec:Shift Compensator}
\ \par
\begin{figure}[t]
    \centering
    \includegraphics[width=0.85\linewidth]{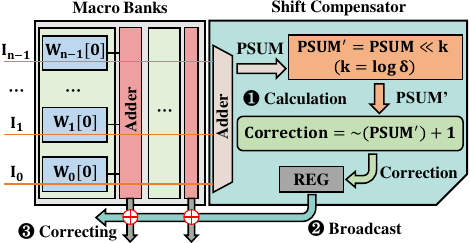}
    \caption{Shift compensator design\revision{: supporting correction term calculation, broadcast and correcting}} 
    \label{fig:SC}
\end{figure}

Alleviating the computational error caused by \wds\ while ensuring the correction step stays off the critical path is central to the design of the shift compensator (SC). Figure~\ref{fig:SC} presents its design, which is strategically located next to PIM macro Banks, sharing input streams to optimize resource usage. The SC performs three main operations: \myding{1} Correction calculation: This step involves summing the inputs, multiplying the sum by $\delta$, and inverting the result. Since $\delta$ is a power of 2, multiplication simplifies to a bit-shifting operation, significantly reducing computational cost. \myding{2} Correction term broadcast: All banks within a Macro share the same input streams and $\delta$ values, allowing a single correction term to be broadcast across all banks. This shared design minimizes hardware redundancy and simplifies the design. \myding{3} Pipelined correcting: To ensure the correction process does not interfere with the critical path, a register is inserted after the correction term calculation. This allows the MAC operation to proceed concurrently with the correction term computation. In the next cycle, the correction term is applied to the MAC output via a pipelined binary addition.

\subsection{IR-Booster Design} \label{sec:booster}
While \lhr\ and \wds\ effectively reduce the \HR\ of neural network weights using quantization algorithms and data distribution techniques, achieving comprehensive architecture-level IR-drop mitigation requires a synergistic software-guided hardware approach. Two primary challenges necessitate this integration. 1) Dynamic mitigation requirements: The available architecture-level mitigation space varies significantly across different neural networks and even operators within the same network. The effectiveness of \lhr\ and \wds\ also differs across workloads. The hardware must dynamically adapt to provide optimal mitigation levels. 2) Input-determined operators: For operators where in-memory data and input data both depend on runtime processing (e.g., QK\textsuperscript{T} and SV in attention), weight-based optimizations such as \lhr\ and \wds\ are insufficient. It highlights real-time IR-drop monitoring and dynamic hardware control for IR-drop mitigation.

To address these challenges, we propose \booster. \revision{Leveraging hardware-based dynamic adjustments and a safety monitoring mechanism, \booster\ allows macros to operate at lower voltages and higher frequencies, thus enhancing energy efficiency and chip performance. Specifically, \booster\ first reserve voltage-frequency pairs for different \tog\ gradients, ensuring safe operation across corresponding IR-drop intensities for Macro Groups during hardware design signoff. Then, during compilation, \booster\ gathers the weight \HR\ information for each Macro after task mapping is completed. Since \HR\ represents the upper bound of \tog, the safe voltage and frequency for each task are determined during pre-processing. At runtime, \booster\ can aggressively adjust the voltage and frequency, while maintaining chip reliability through a hardware-based IRFailure detection and recalculation mechanism.}
\subsubsection{IR-Booster V-f Pair Adjustment}\label{subsubsec:v-f pair}\ \par 
As illustrated in Figure~\ref{fig:booster}-(a), \revision{due to hardware design requirements, multiple PIM macros are integrated into a \textbf{Macro Group}, sharing a unified power supply and operating at the same frequency.} For each Macro Group, \booster\ provides a selection of voltage-frequency pairs, referred to as \textbf{V-f pairs}, each associated with a specific \tog\ value. These V-f pairs define operational conditions under which the macro Group can safely function without exceeding the designated \tog. Hardware engineers validate these pairs when the Macro Group is issued as an independent IP (intellectual property), ensuring reliability for the specified \tog\ levels.

\begin{table}[t]
\centering
\caption{Safe level and corresponding initialized a-level (\%)}
\label{tab:aggressive Gear}
\resizebox{\columnwidth}{!}{%
\begin{tabular}{|c|c|c|c|c|c|c|c|c|c|c|}
\hline
safe level      & 100 & 60 & 55 & 50 & 45 & 40 & 35 & 30 & 25 & 20 \\ \hline
a-level$_0$         & 60  & 40 & 35 & 35 & 35 & 30 & 30 & 25 & 20 & 20 \\ \hline
\end{tabular}%
}
\end{table}

\begin{figure}[t]
    \centering
    \includegraphics[width=0.9\linewidth]{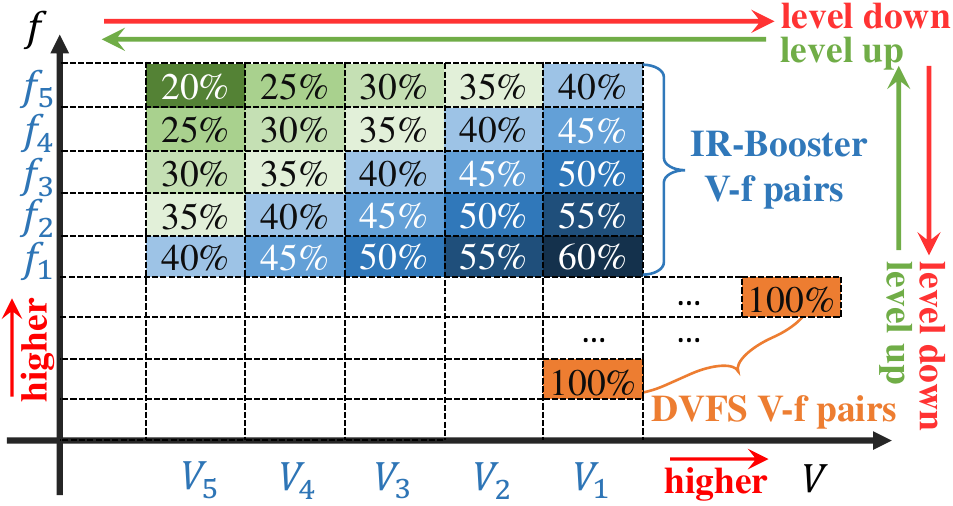}
    \caption{\booster\ and DVFS V-f pairs\revision{: level up of V-f pairs means lower voltage or higher frequency or both of them}} 
    \label{fig:Gears}
\end{figure}

\booster\ can be regarded as an extension of Dynamic Voltage and Frequency Scaling (DVFS) technology. Figure~\ref{fig:Gears} illustrates the main differences between \booster\ pairs and traditional DVFS technology (noting that the figure depicts trends rather than linear relationships in value): DVFS ensures the chip reliability by still signing off V-f pairs with the worst-case IR-drop conditions (that is \tog=100\%)~\cite{dvfs0,dvfs-intel,dvfs1,dvfs2}. Therefore, DVFS could only increase/decrease voltage and frequency simultaneously based on the urgency of the processor's current tasks. In contrast, \booster\ leverages the architecture-level IR-drop margin to enable more flexible and granular adjustments. That is, \booster\ can reduce voltage while maintaining the operating frequency or increase frequency while holding the power supply conditions constant.

A series of V-f pairs corresponding to the same \tog\ value form a subset, and their corresponding \tog\ value is recorded as \textbf{level}. \revision{Based on workload profiling, we first define the V-f pair level range to be 20\%\thicktilde60\%. Sensitivity analysis indicates that narrowing this range by 5\% results in over 17\% loss in IR-drop mitigation capability, while widening the range has minimal improvement (<3\%). Additionally, using a step of 6\% (4x4 V-f pairs) or larger leads to more than 8\% loss in capability, as fewer V-f pairs affect fine-grained control. While a step below 5\% can yield better IR-drop mitigation (\thicktilde6\%), increasing the number of V-f pairs to 36 or more significantly raises hardware design complexity and cost, which is unacceptable. Therefore, the V-f pair level step is set to 5\%.}

For a given operator, \booster\ first selects a \textbf{safe level} that can ensure reliable operation based on \HR\ optimization results from \lhr\ and \wds. Then, \booster\ determines a more performance- or energy-efficient \textbf{aggressive level} according to the safe level and adjusts it based on the feedback from the hardware IR Monitor.

\begin{algorithm}[t]
    \begin{algorithmic}[1]
        \STATE{\textbf{initialize: }$SafeCounter = 0, a\text{-}Level = a\text{-}Level_0$}
        \STATE{$Level = a\text{-}Level$}  \textcolor{myblue}{\textit{/*(Start with a-Level)*/}}
        \FOR{Each Cycle}
        \IF{$IR Failure==True$}
            \STATE{$Level = SafeLevel$} \textcolor{red}{\textit{/*(Set Safe Level)*/}}
            \IF{$SafeCounter < 0.2\beta$}
            \STATE{\textbf{set: }$SafeCounter=0$}
                \STATE{$a\text{-}Level = Down(a\text{-}Level)$} \textcolor{red}{\textit{/*(a-Level Down)*/}}
            \ENDIF
            \STATE{\textbf{set: }$SafeCounter=0$}
            
        \ELSIF{Macro frequency adjustment in same Set}
            \STATE{$Level = Set\text{-}Level$ \textit{/*(Frequency Synchronization)*/}}
            \STATE{\textbf{set: }$SafeCounter=0$}
        \ELSE
            \STATE $SafeCounter++$
            \IF{$SafeCounter==\beta$}
                \STATE{$Level = a\text{-}Level$}  \textcolor{myblue}{\textit{/*(Back to a-Level)*/}}
            \ENDIF
            \IF{$SafeCounter > 2\beta$}
                \STATE{$a\text{-}Level = UP(a\text{-}Level)$} \textcolor{OliveGreen}{\textit{/*(a-Level Up)*/}}
                \STATE{$Level = a\text{-}Level$}
                \STATE{\textbf{set: }$SafeCounter=\beta$}
            \ENDIF
        \ENDIF
        \ENDFOR
    \end{algorithmic}
    \caption{\booster\ level adjustment for each Macro Group}
    \label{algo:dvfs}
\end{algorithm}
\noindent \textbf{Software-guided safe level selection}:
For operators like conv, Q/K/V generation, and linear layers, where their weights are the in-memory data, the \HR\ of each macro and the worst \HR\ in each Group (HR\textsubscript{G}) can be pre-determined after task mapping. Using HR\textsubscript{G}, \booster\ selects the nearest higher \tog\ level (rounded to the nearest 5\%) as the safe level. For example, if HR\textsubscript{G} is 47.5\%, the safe level is set to 50\%, and the optional V-f pairs are $\{V_3\text{-}f_1,V_4\text{-}f_2,V_5\text{-}f_3\}$. The safe level of Groups with $\text{HR\textsubscript{G}}>60\%$ revert to DVFS (100\% \tog), which is quite rare after \lhr\ and \wds\ optimization. For input-determined operators such as QK\textsuperscript{T} and SV in transformer attention, both in-memory data and input data are generated from preceding calculations. Their \HR\ values cannot be offline pre-determined by the algorithm, and thus, these operators default to a 100\% safe level. 

\noindent \textbf{IRFailure-aware aggressive level adjustment}:
After selecting a safe level, \booster\ aims to operate at an aggressive level (a-level) that provides higher performance or energy efficiency. This process uses real-time feedback from the IR Monitor to iteratively refine the a-level, as outlined in Algorithm~\ref{algo:dvfs}.
IRFailure-aware aggressive level adjustment mainly consists of three parts: 
1) Initialization: \booster\ initializes with an a-level\textsubscript{0} derived from the pre-determined safe level (lines 1-2). The values for a-level\textsubscript{0}, shown in Table~\ref{tab:aggressive Gear}, are determined based on profiling results. This aligns with the intuition that a \revision{higher} safe level provides greater remaining utilization space for optimization.
2) IRFailure-driven level adjustment: Whenever IR-drop violation occurs, the level is set to the safe level to ensure the chip reliability (lines 4-5). If the failure interval is too short (less than $0.2\beta$), it indicates that the a-level is overly aggressive, prompting \booster\ to level down (lines 6-9). 
3) A-level optimization: If the Macro Group operates without failures for continuous $\beta$ cycles, the a-level is restored (lines 13-14). If safe operation extends beyond $2\beta$ cycles, the a-level is increased by 5\% (level up), unlocking additional performance or power savings (lines 16-19).
            

\noindent \textbf{Sprint mode and low-power mode}:
After determining the \tog\ level, \booster\ selects an appropriate V-f pair within the subset of this level. \booster\ offers two modes of operation. In sprint mode, \booster\ prioritizes high-voltage, high-frequency combinations to maximize throughput. Conversely, when external operating conditions impose constraints (e.g., limitations on heat generation or power consumption rates), \booster\ operates in low-power mode, favoring low-voltage, low-frequency pairs.

\revision{These two modes are designed to simplify user configuration and ensure clearer experimental results. However, the design of \booster\ fully supports flexible selection of V-f pairs. When using \name, users can further customize the \booster\ strategy through the available interfaces to tailor it for specific needs.}

\subsubsection{IRFailure-driven recomputing mechanism}\ \par
To ensure the chip's reliability, detecting IRFailures and performing recomputing after such errors are crucial.

\begin{figure}[t]
    \centering
    \includegraphics[width=0.9\linewidth]{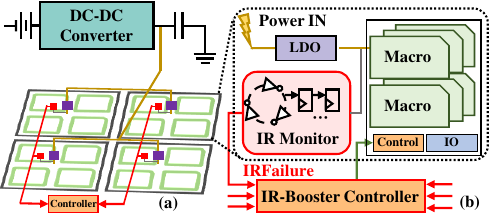}
    \caption{\revision{(a) Power supply and V-f adjustment at Macro Group granularity (b) \booster\ regulated by IRFailure}}
    \label{fig:booster}
\end{figure}

\noindent \textbf{IR Monitor}: 
Building on the design proposed in~\cite{IR-Monitor}, we introduce a simplified voltage monitoring device called the IR monitor. Each IR monitor comprises multiple inverter (INV) gates configured as a free-oscillating loop, functioning as a Voltage-Controlled Oscillator (VCO). The IR-drop reduces the supply voltage, which in turn affects the VCO's oscillation frequency, enabling a voltage-to-frequency conversion. By sampling and analyzing the VCO's phase, the device generates a corresponding digital signal. If the supply voltage drops below a predefined threshold, the monitor triggers an ``IRFailure'' signal. As shown in Figure~\ref{fig:booster}-(b), IR monitors are strategically embedded between Macro Groups and their corresponding Low Dropout Regulators (LDOs) to monitor the voltage supplied to all Groups. The “IRFailure” signals from these monitors are sent to the Booster Controller, which oversees the adjustment of voltage-frequency (V-f) pairs and manages recomputing.

\begin{figure}[t]
    \centering
    \includegraphics[width=0.95\linewidth]{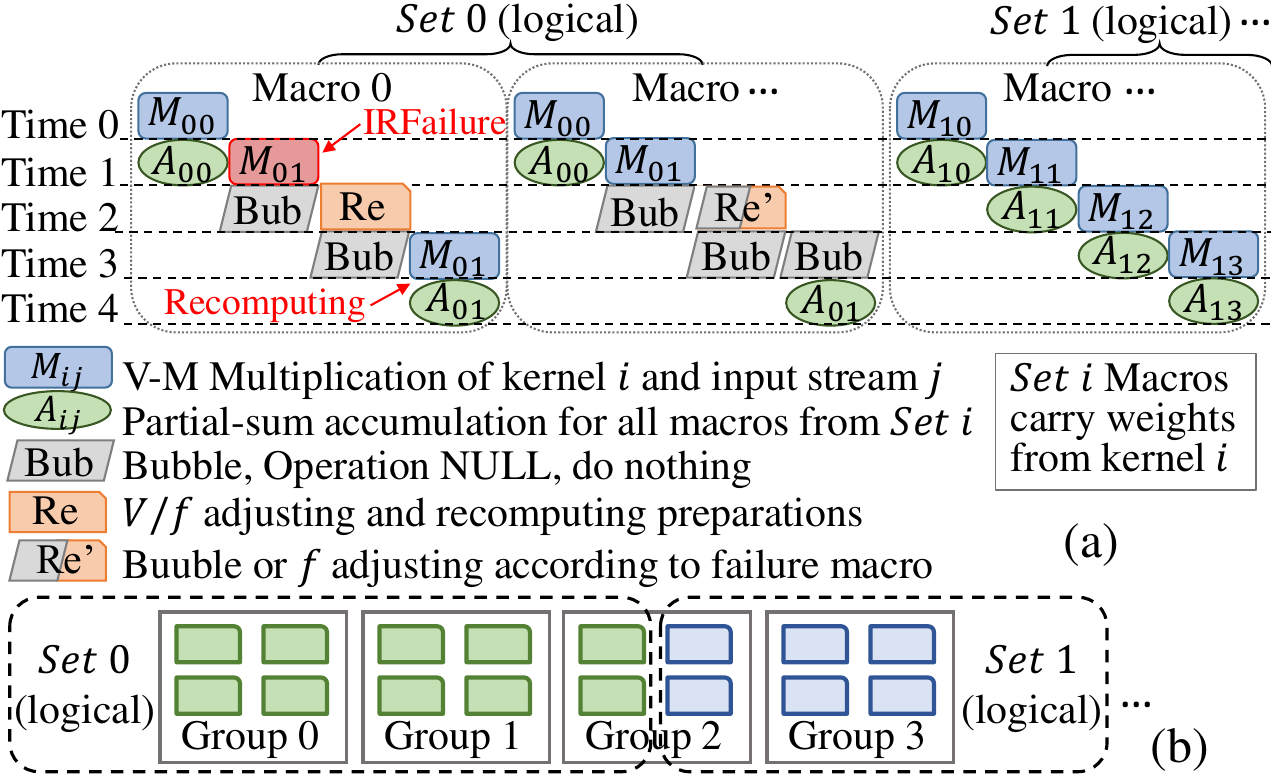}
    \caption{(a) \booster\ recomputing: minimizing the delay due to interference between macros \revision{(b) Macro Set definition}} 
    \label{fig:pipe}
\end{figure}

\noindent \textbf{Booster Controller}: 
The Booster Controller processes IRFailure signals from all Groups, directing the affected Groups to suspend computation, adjust their V-f pairs, and initiate recomputing. \revision{As shown in the Figure~\ref{fig:pipe}-(b), task allocation is performed at the macro granularity. During inference, multiple macros from different groups are combined to compute an operator, forming a logical Macro \textbf{Set}.} Consequently, while macros within the same physical Group share identical frequency and voltage settings, all macros within a logical Set must operate at the same frequency to ensure consistency, even if their voltages could differ.
When any macro requires recomputing due to an IRFailure, other macros within the same Set temporarily stall the pipeline and store partial sums until the recomputing completes, ensuring data consistency. Importantly, this process does not disrupt macros in other Sets. For example, as depicted in Figure~\ref{fig:pipe}-(a), when an IRFailure is detected in macro 0 of Set 0, the Booster Controller halts the operations of all macros in Set 0. It then instructs macro 0 to adjust its voltage and frequency settings before performing the necessary recomputing. Priority is given to increasing the voltage to maintain the current frequency whenever possible. Other macros in Set 0 do not perform additional computations, thereby minimizing power consumption. 

\subsection{HR-aware Task Mapping} \label{sec:Mapping}
Increasingly complex application scenarios and models often require multiple operators to run concurrently, significantly complicating task mapping. For instance, UniAD~\cite{UniAD}, BEVFormer~\cite{BEVFormer}, and TransFuse~\cite{transfuse}—which combine conv and transformer layers with highly varying \HR\ values—are widely employed in visual perception, autonomous driving, and medical image segmentation, respectively. Even within a single transformer-based model, the \HR\ and corresponding safe levels vary significantly across layers, such as Q/K/V generation, QK\textsuperscript{T}, SV, and linear layers.

Previous operator scheduling and mapping methods~\cite{mapping0, mapping1, mapping2} generally assign split operators to the available computing power of a region sequentially after considering operator scheduling and mapping. This approach works well in traditional PIM architectures, where macros operate under uniform conditions with the same fixed V-f pair. However, in PIM chips supported by \booster, once a weight is assigned, each macro has its own independent \HR\ and, consequently, a distinct safe V-f level. Since the PIM chip adjusts the power supply and V-f at the Macro Group level, the macros within a group are constrained by the macro requiring the lowest V-f level. Additionally, macros handling tasks from the same operator must operate at the same frequency. These constraints make it inefficient to assign split operators to PIM macros using sequential or random mapping strategies, resulting in suboptimal performance. To address this, we propose HR-aware task mapping, which is integrated into the compiler to optimize task-to-macro mapping after operator scheduling and before inference begins.


Exploring all possible task mapping combinations for a PIM chip with 64 macros during compilation is computationally infeasible. Additionally, \revision{an absolutely optimal mapping does not exist because of unknown input features}. Therefore, we implement Algorithm~\ref{algo:mapping} based on simulated annealing~\cite{SA}. Key parameters include a temperature reduction coefficient $q$ of 0.95, an initial normalized temperature $T_0$ of 1, and an iteration limit of 500. We adopt the normalized exponential (NE) acceptor function to minimize the influence of system-level parameter adjustments. It terminates early if ten consecutive attempts are rejected.

\begin{algorithm}[t]
    \begin{algorithmic}[1]
        \STATE{\textbf{initialize: }$M_{best} \leftarrow M_0, T\leftarrow T_0$}
        \STATE{$S_{best} \leftarrow S_0\leftarrow Score(M_0)$}
        \FOR{$i \leftarrow 0 \text{ to } Steps$}
            \STATE{$T \leftarrow qT$, $M_{new}=Switch(M),\ S_{new}=Score({M_{new})}$}
            \STATE{$\Delta S=S_{new}-S$}
            \IF{$\Delta S<0\ \textbf{or}\ Random()<\exp({\frac{-\Delta S}{0.5S_0T}})$}
                \IF{$S_{new}<S_{best}$}
                    \STATE{$M_{best}\leftarrow M_{new}, S_{best}\leftarrow S_{new}$}
                \ENDIF
                \STATE{$M\leftarrow M_{new}, S\leftarrow S_{new}$}
            \ENDIF
            \IF{$End\text{-}Condition()$}
                \STATE{\textbf{break}}
            \ENDIF
        \ENDFOR
        \RETURN $M_{best}$
    \end{algorithmic}
    \caption{\HR-aware Task Mapping Algorithm}
    \label{algo:mapping}
\end{algorithm}
The mapping transition function randomly selects two macros and their corresponding tasks from different Groups and exchanges them. Importantly, an ``empty macro'' option is introduced, allowing one or two macros to remain unmapped in extreme cases. This prevents interference between tasks from different operators with significantly different \HR\ values. The mapping evaluation function relies on a lightweight simulator. Based on profiling input data, the simulator generates a 100-step input flip sequence sampled from a normal distribution, which is then combined with the \HR\ values assigned to each macro. The simulator uses this data to estimate the end-to-end delay and power consumption for the given mapping. If a macro remains vacant in the final mapping plan, it indicates that significant \HR\ disparities prevented tasks from being mapped together, effectively avoiding interference. In such cases, tasks already assigned to the Group can be redistributed evenly across all macros within the Group. It balances the load and reduces the Group's safe level, further optimizing performance.

\section{Evaluation} \label{sec:evaluation}

\subsection{Evaluation Setup} \label{subsec:setup}

We evaluate the effectiveness of \name\ using a combination of software experiment results and hardware post-layout simulation~\cite{TOOL_redhawk-sem}. \revision{Specifically, the HR results (Sections \ref{subsec:HR and Acc}\thicktilde\ref{subsec:sparse}) are derived from our profiling of the adjusted network weights, while the IR-drop and PPA data (Sections \ref{subsec:hardware results}\thicktilde\ref{subsec:overhead}) are obtained from the post-layout simulations, a widely trusted method in hardware design.} We selected ResNet18~\cite{ResNet}, MobileNetV2~\cite{mobilenets}, and YOLOv5~\cite{yolov5} as conv-based networks, and Llama3.2-1B~\cite{Llama3}, ViT~\cite{ViT}, and GPT2~\cite{GPT2} as transformer-based networks. 
YOLOv5 uses COCO~\cite{COCO}, Llama3.2 and GPT2 utilize Wikitext2~\cite{wikitext}, and the remaining networks rely on ImageNet (ILSVRC 2012)~\cite{imagenet}. The hardware architecture is a chip design featuring two RISC-V cores and 16 macro groups. Each group contains four macros. \revision{The architecture parameters are abstracted from a standard half-height, half-length PCIe acceleration card based on SRAM PIM with support for 16-channel FHD encoding, and a maximum computing power of 256 TOPS.}

\begin{table}[t]
\centering
\caption{HR\textsubscript{average} and HR\textsubscript{max} reduction over baseline\cite{quant-baseline}}
\label{tab:HR reduction}
\resizebox{0.99\columnwidth}{!}{%
\begin{tabular}{@{}|cc|c|c|c|c|c|c|@{}}
\toprule
\multicolumn{2}{|c|}{\textbf{Model}}                                                & \textbf{ResNet18}  & \textbf{MobileNet}  & \textbf{YOLOv5} & \textbf{ViT} & \textbf{Llama3} & \textbf{GPT2}\\ \midrule
\multicolumn{1}{|c|}{\multirow{3}{*}{\textbf{HR\textsubscript{aver}}}}  &+\lhr             & 28\%               & 29\%               & 23\%             & 25.9\%       & 25.9\%          & 30.7\%       \\
\multicolumn{1}{|c|}{}                                          & +\wds($\delta$=8)  & 39\%               & 30.6\%             & 31.5\%           & 31.9\%       & 30.7\%          & 38\%         \\
\multicolumn{1}{|c|}{}                                          & +\wds($\delta$=16) & 45.6\%             & 33.6\%             & 38.6\%           & 35.6\%       & 36.3\%          & 41.5\%       \\ \midrule
\multicolumn{1}{|c|}{\multirow{3}{*}{\textbf{HR\textsubscript{max}}}}      &+\lhr             & 28.6\%             & 30.5\%             & 24.3\%           & 26.8\%       & 26.5\%          & 31.2\%       \\
\multicolumn{1}{|c|}{}                                          & +\wds($\delta$=8)  & 37.5\%             & 32.2\%             & 30.7\%           & 32.2\%       & 32\%            & 37.5\%       \\
\multicolumn{1}{|c|}{}                                          & +\wds($\delta$=16) & 44.5\%             & 35.1\%             & 37.6\%           & 36.2\%       & 38.3\%          & 41.2\%       \\ \bottomrule
\end{tabular}%
}

\end{table}

\begin{figure}[t]

    \centering
    \includegraphics[width=0.98\linewidth]{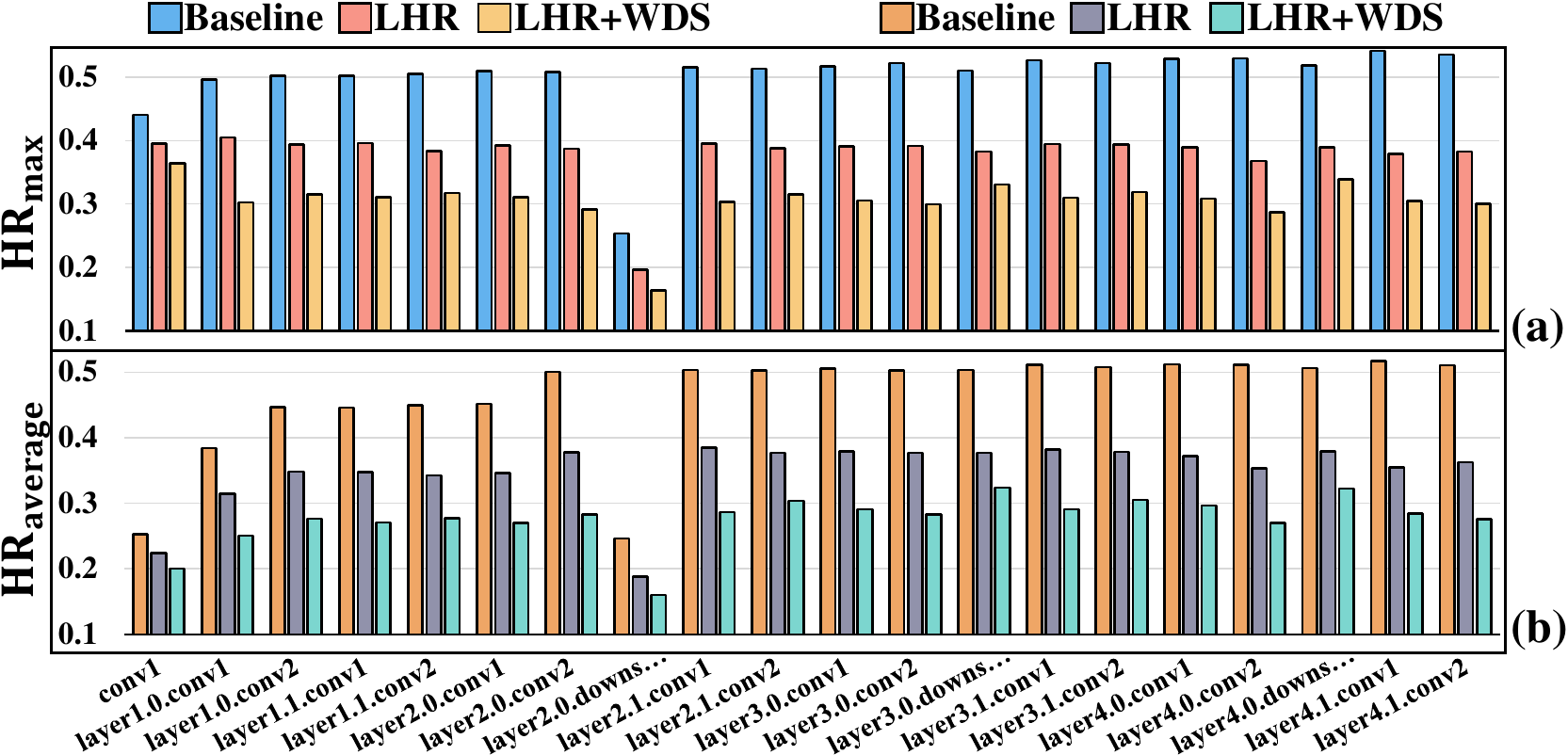}
    \caption{HR reduction of each layer in ResNet18~\cite{ResNet}} 
    \label{fig:layers}
\end{figure}

\subsection{HR Decrease and Accuracy Impact} \label{subsec:HR and Acc}

Table~\ref{tab:HR reduction} demonstrates that both \lhr\ and \wds\ significantly reduce HR\textsubscript{max} and HR\textsubscript{average} compared to the baseline~\cite{quant-baseline}, a widely adopted quantization-aware-training (QAT) algorithm. ``+LHR'' means adding the \lhr\ regularization term on the basis of baseline, and ``+WDS($\delta$)'' means adding WDS on the basis of ``+LHR''. 

Figure~\ref{fig:layers} provides additional insights by illustrating how \lhr\ and \wds\ ($\delta = 16$) impact HR across all layers of ResNet18. While initial layers with smaller kernels show imbalanced HR due to under-utilized macros, most layers exhibit similar HR values. This suggests a uniform \HR\ distribution within individual layers and structural homogeneity across similar layers in the network. These findings strongly support our HR-aware task mapping design.

\begin{figure}[t]
    \includegraphics[width=0.98\linewidth]{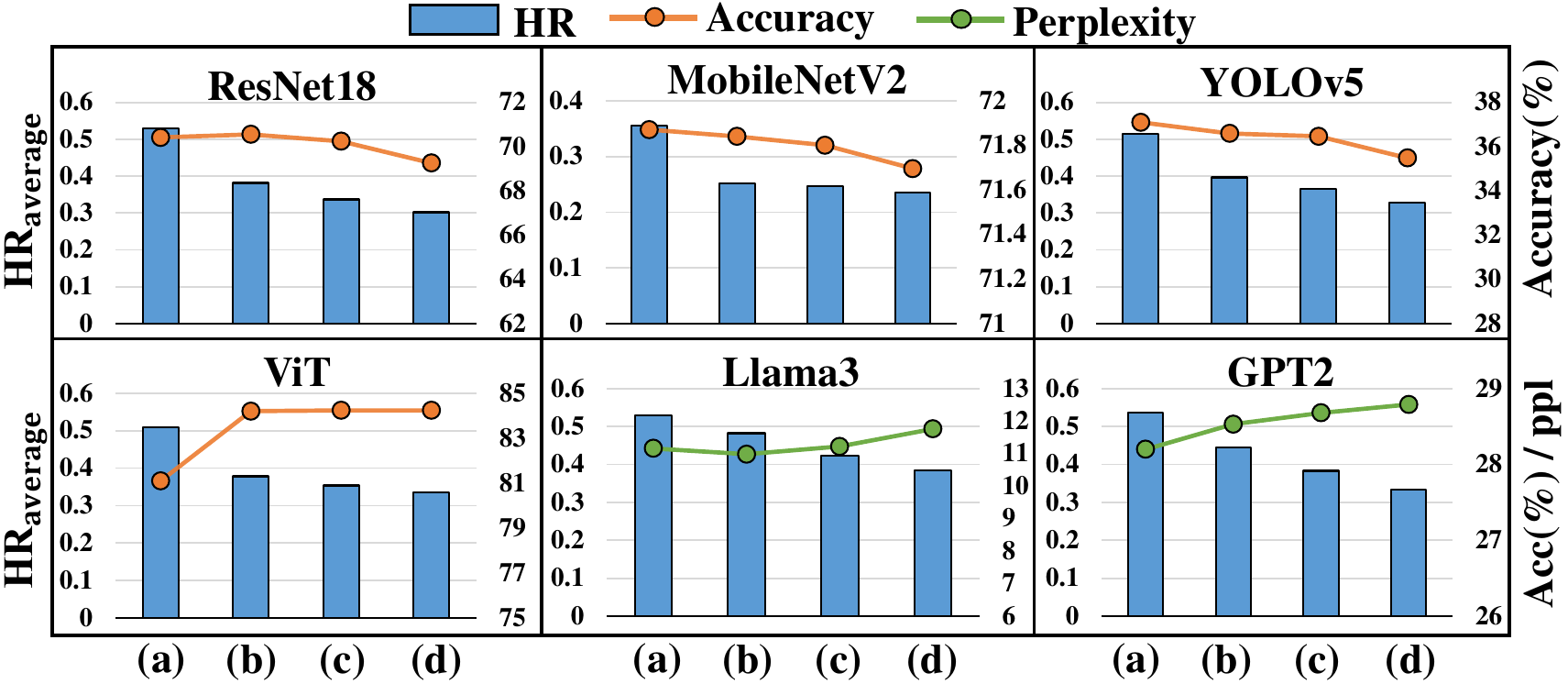}
    \caption{\HR\ Decreasing and Accuracy Influence for: (a) Baseline~\cite{quant-baseline} (b) +\lhr\ (c) +\wds\ ($(\delta=8)$) (d) +\wds\ ($\delta=16$)} 
    \label{fig:HRandAcc}

\end{figure}

Figure~\ref{fig:HRandAcc} highlights the trade-off between HR reduction and accuracy across different workloads. 
Applying \lhr\ and \wds\ yields substantial HR reductions with minimal accuracy degradation. \revision{ViT and Llama3 even show improved accuracy. This is because moderate quantization does not notably affect accuracy but enhances the network’s generalization ability. Similar findings have been observed in previous quantization works, such as~\cite{QAT-0,QAT-1}.}

\begin{table}[t]
\centering
\caption{HR\textsubscript{average} and accuracy impact on PTQs with \lhr}
\label{tab:ptq}
\resizebox{0.99\columnwidth}{!}{%
\begin{tabular}{@{}|c|cccc|cccc|@{}}
\toprule
\textbf{PTQ}     & \multicolumn{4}{c|}{\textbf{OmniQuant~\cite{round-OmniQuant}}}                                                                                       & \multicolumn{4}{c|}{\textbf{BRECQ~\cite{brecq}}}                                                                                           \\ \midrule
Model            & \multicolumn{2}{c|}{GPT2}                                               & \multicolumn{2}{c|}{Llama3.2-1B}                    & \multicolumn{2}{c|}{ResNet18}                                           & \multicolumn{2}{c|}{MobileNetv2}                    \\ \midrule
Metric           & \multicolumn{1}{c|}{HR\textsubscript{aver}} & \multicolumn{1}{c|}{ppl}   & \multicolumn{1}{c|}{HR\textsubscript{aver}} & ppl    & \multicolumn{1}{c|}{HR\textsubscript{aver}} & \multicolumn{1}{c|}{acc (\%)}   & \multicolumn{1}{c|}{HR\textsubscript{aver}} & acc (\%)    \\ \midrule
\textbf{w/o LHR} & \multicolumn{1}{c|}{0.51}                  & \multicolumn{1}{c|}{28.69} & \multicolumn{1}{c|}{0.53}                  & 11.16  & \multicolumn{1}{c|}{0.5}                   & \multicolumn{1}{c|}{73.02}  & \multicolumn{1}{c|}{0.49}                  & 69.715 \\ \midrule
\textbf{w LHR}   & \multicolumn{1}{c|}{0.47}                  & \multicolumn{1}{c|}{28.72} & \multicolumn{1}{c|}{0.49}                  & 10.947 & \multicolumn{1}{c|}{0.47}                  & \multicolumn{1}{c|}{72.9} & \multicolumn{1}{c|}{0.46}                  & 69.71  \\ \bottomrule
\end{tabular}%
}

\end{table}

\subsection{\lhr\ Combination with PTQ Methods}

We evaluated the integration post-training-quantization (PTQ) methods with \lhr\ in Table~\ref{tab:ptq}, specifically OmniQuant~\cite{round-OmniQuant} for LLMs and BRECQ~\cite{brecq} for traditional convolutional networks. Since PTQ methods prioritize rapid, resource-constrained quantization and do not involve adjustments or joint training of the original model, their ability to fine-tune weights is limited. As a result, the HR reduction achieved by combining \lhr\ with PTQ methods is slightly less pronounced compared to QAT~\cite{quant-baseline}. Nevertheless, incorporating \lhr\ into PTQ still significantly reduces HR with minimal accuracy loss. This seamless integration demonstrates \lhr's ability to enhance HR performance when combined with various PTQ methods.

\subsection{Configuration of $\delta$ for \wds}

\begin{figure}[t]
    \centering
    \includegraphics[width=0.95\linewidth]{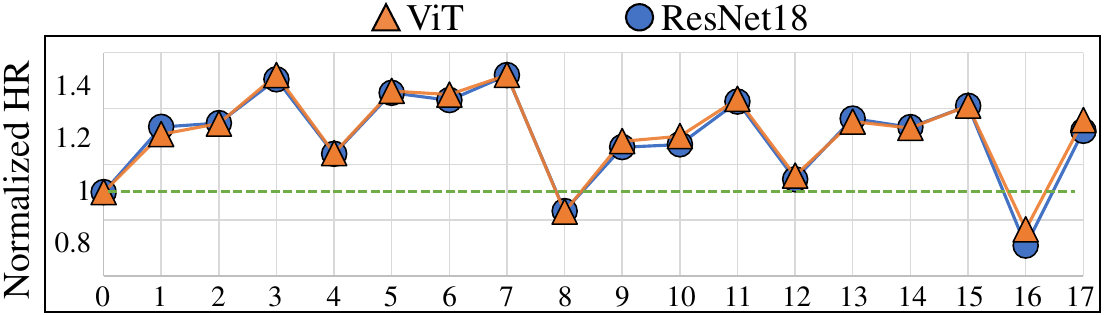}
    \caption{Impact of different \revision{$\delta$} on \wds} 
    \label{fig:delta}
\end{figure}
Figure~\ref{fig:delta} illustrates the impact of the $\delta$ selection in the \wds\ technique on the overall HR of the network, with data normalized to the HR of weights quantized using \lhr. Across different workloads, the trend of HR reduction with varying $\delta$ values remains consistent, demonstrating the broad applicability of \wds. Aligning with our analysis in Section~\ref{sec:wds}, a precise configuration of $\delta$ is crucial for achieving HR reduction: for 8-bit quantized weights, only $\delta$ values of 8 or 16 effectively reduce HR. Other configurations result in weight distributions that align with higher HR regions, leading to negative effects. This specific choice of $\delta$ is governed by the binary 2's complement representation method. The fact that $\delta$ must be a power of 2 is the basis for using the shift method in the shift compensator to simplify multiplication calculations.

\subsection{\revision{Comparison and Combination with Pruning}} \label{subsec:sparse}
\begin{figure}[t]
    \centering
    \includegraphics[width=0.99\linewidth]{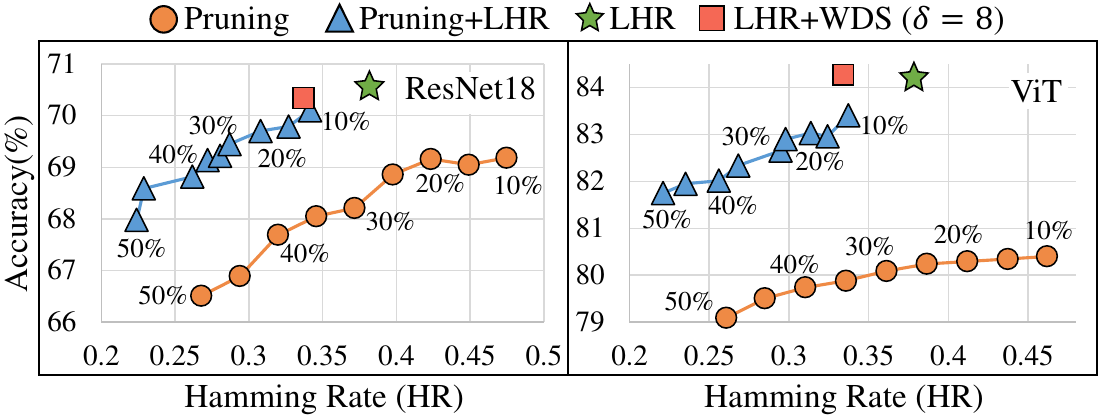}
    \caption{\revision{Comparison and combination of \lhr\&\wds\ and pruning on ResNet18 and ViT with different sparse target}} 
    \label{fig:sparse}
\end{figure}

\revision{Figure~\ref{fig:sparse} illustrates the comparison and relationship between our methods and pruning. For the comparison, we employ the open-source SparseML framework alongside the Gradual Magnitude Pruning*~\cite{gmp} to generate networks with varying sparsity levels.}

\revision{Network pruning can reduce HR, making it a natural candidate for integration with \booster\ to mitigate IR-drop. However, we introduce \lhr\ and \wds\ for two key advantages. First, \lhr\ serves as a regularization term during quantization, ensuring that weight values are not drastically altered, which helps maximize inference accuracy. Second, \lhr\ and \wds\ are orthogonal to pruning, allowing them to be seamlessly combined. When integrated, they can further reduce HR at the cost of a slight accuracy degradation.}

\begin{figure}[b]
    \centering
    \includegraphics[width=\linewidth]{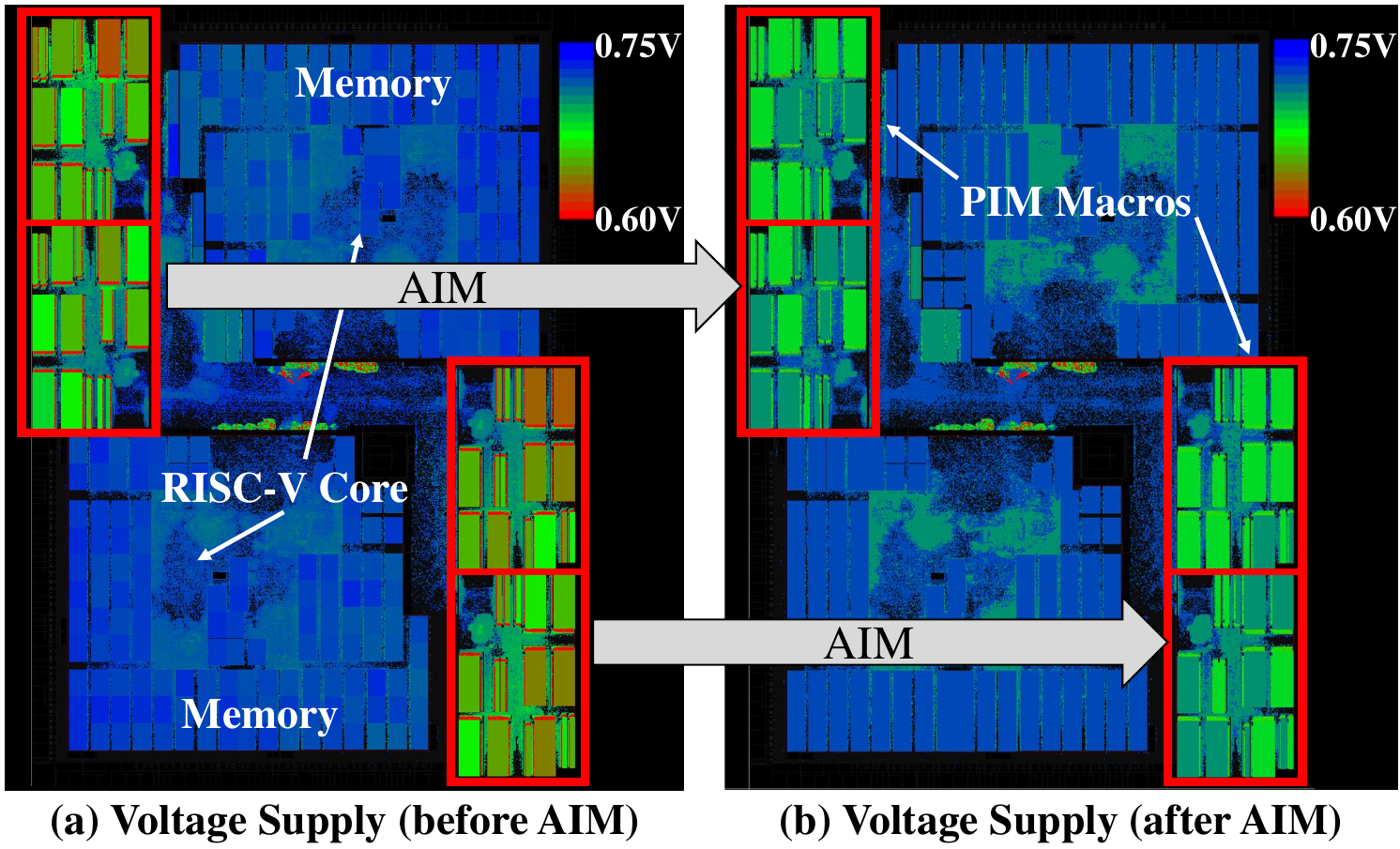}
    \caption{IR-drop mitigation in a 7nm 256-TOPS PIM \revision{layout}} 
    \label{fig:IR Mitigation}
\end{figure}

\begin{figure}[t]
    \centering
    \includegraphics[width=\linewidth]{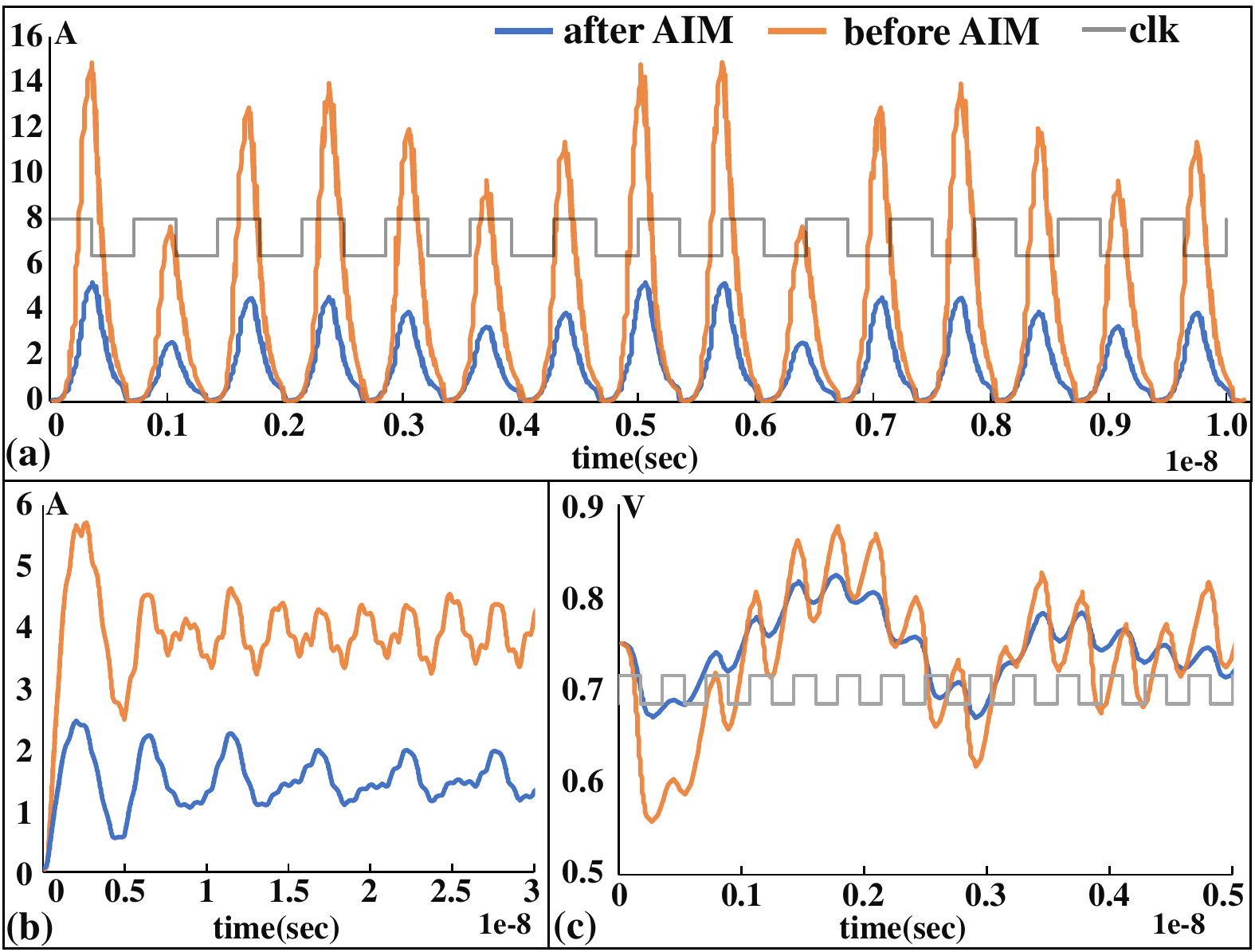}
    \caption{(a) Demanded drive current, (b) Bump voltage, and (c) Bump current before/after~\name} 
    \label{fig:hardware result}
\end{figure}

Utilizing post-layout simulation~\cite{TOOL_redhawk-sem}, we evaluate \name's performance \revision{based} on our 7nm 256-TOPS PIM \revision{design}. Figure~\ref{fig:hardware result}-(a) shows the reduction in demand drive current across clock cycles, while Figures~\ref{fig:hardware result}-(b) and (c) illustrate \name's influence on bump current and voltage. ``bump'' refers to small metal protrusions utilized to establish connections between the chip and its external interfaces, which are essential for reliable chip packaging and interconnection. These metrics are critical for chip design. Notably, \name\ achieves reductions in both demanded drive current and bump current, along with a stabilization of bump voltage. This outcome reflects effective IR-drop mitigation and indicates enhanced chip performance. Furthermore, the decrease in current directly reduces power consumption and heat generation, significantly improving the chip's energy efficiency and thermal management.

\subsection{Hardware Results on 256-TOPS PIM \revision{Design}} \label{subsec:hardware results}
Figure~\ref{fig:IR Mitigation} illustrates the distribution of IR-drop across the chip layout and the mitigation effects achieved by \name\ on the PIM chip. The IR-drop in the RISC-V core and on-chip memory is relatively minimal, with hotspots predominantly concentrated in the PIM macros. This observation is consistent with our earlier analysis, which identified the macros as the primary contributors to significant IR-drop. While \name's data adjustments and hardware modifications introduce slight variations in the IR-drop of RISC-V core and memories, these impacts are negligible given the consistency of primary IR-drop hotspots in macros.

\revision{According to the post-layout simulation results of} the 7nm PIM chip with an operating voltage of 0.75V, \name\ can reduce the IR-drop from 140mV to 58.1\thicktilde43.2mV within a macro, providing \textbf{58.5\%\thicktilde69.2\%} architecture-level IR-drop mitigation. Additionally, \name\ provides a \textbf{1.91\thicktilde2.29$\times$} improvement in energy efficiency for each macro (4.2978mW$\to$2.243\thicktilde1.876mW) and enhances overall chip performance by {\bf 1.129\thicktilde1.152$\times$} (256TOPS$\to$ 289\thicktilde295TOPS) in low-power mode or sprint mode, respectively.

\subsection{Trade-off in \booster\ and $\beta$ Configuration} \label{subsec:beta}
\begin{figure}[t]
    \centering
    \includegraphics[width=0.98\linewidth]{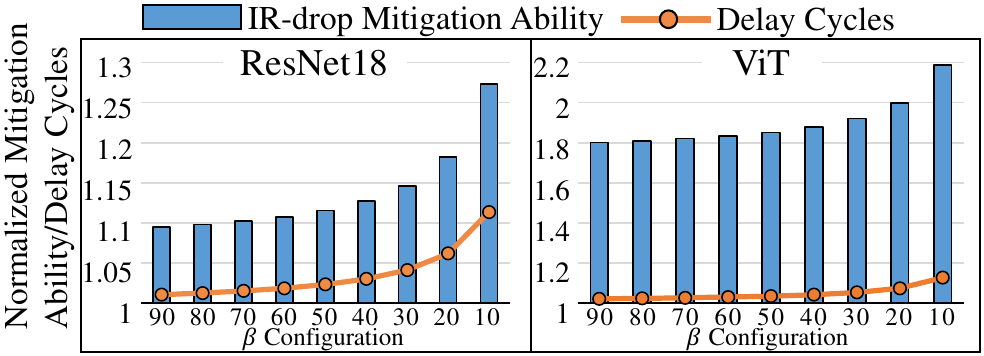}
    \caption{Impact of different $\beta$\revision{: normalized against \booster\ operating without aggressive V-f adjustment}} 
    \label{fig:beta}
\end{figure}

Figure~\ref{fig:beta} shows the impact of various $\beta$ configurations on \booster\ performance. Both IR-drop mitigation ability and delay cycles are normalized against \booster\ operating without aggressive level adjustment (i.e., running exclusively at a safe level). A smaller $\beta$ value corresponds to tighter clock adjustments, which enhances IR-drop mitigation but increases the occurrence of IRFailures. These failures, in turn, lead to additional delay cycles required to complete the workload. As analyzed in Section~\ref{subsec:ablatian}, transformer-based ViT benefits more significantly from \booster's aggressive adjustments compared to convolution-based workloads. This difference stems from ViT's heavier reliance on dynamic, input-dependent operations, which are better supported by \booster’s ability to adaptively mitigate IR-drop.

\subsection{Ablation Study} \label{subsec:ablatian}
\begin{figure}[t]
    \centering
    \includegraphics[width=0.98\linewidth]{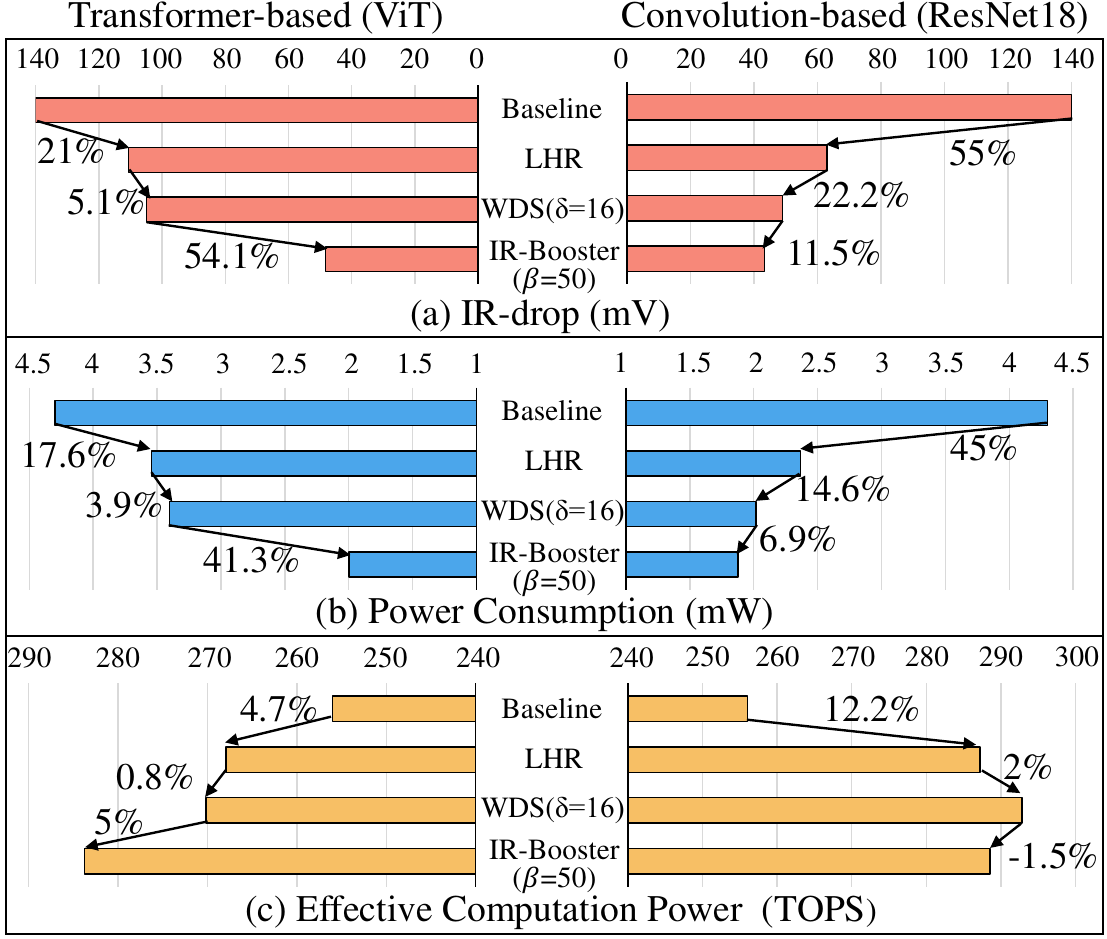}
    \caption{Ablation study in IR-drop, power, and performance} 
    \label{fig:ablation}
\end{figure}

We conducted an ablation experiment to analyze the contribution of each component of \name. It is important to note that using software operations (\lhr, \wds) alone does not directly reduce power consumption or improve performance. Therefore, the data represented by \lhr\ and \wds\ corresponds to the results obtained with basic \booster\ support, using a safe-level configuration, to reflect the impact of the software methods. We selected two representative workloads: ViT for transformer-based networks and ResNet18 for convolution-based networks. Although the effects of \name\ vary across different workloads, they follow similar trends. Figure~\ref{fig:ablation}-(a) demonstrates the IR-drop mitigation effects, while Figure~\ref{fig:ablation}-(b) and (c) illustrate the improvements in energy efficiency and processing performance under low-power and sprint modes, respectively. 

To isolate the effect of PIM macros from the RISC-V core and other memory components, we evaluate energy efficiency by measuring the average power consumption (in mW) of each macro. Additionally, to assess potential degradation in effective computing power due to IRFailures—i.e., scenarios in which a macro’s failure disrupts the operation of others—we use overall processing performance (in TOPS) as the key metric. This methodology offers a comprehensive understanding of how \name\ improves both energy efficiency and computational performance, ensuring that IR-drop mitigation does not compromise system reliability or throughput.

As illustrated in Figure~\ref{fig:ablation}, under transformer-based workloads, the primary improvements stem from \booster, whereas for convolution-based workloads, \lhr\ provides the main benefits. This divergence arises from the inherent characteristics of the workloads: For transformer-based workloads, Q, K, and V matrices are all related to user input, preventing effective pre-profiling by software and thus requiring hardware-based IR-drop mitigation. In contrast, conv operators enable pre-optimization and profiling of weights, so software-guided methods can effectively mitigate IR-drop.

Figure~\ref{fig:ablation}-(c) also reveals a limitation of \booster\ in enhancing the chip’s effective performance.  Its aggressive voltage adjustment strategy can sometimes lead to IRFailures, requiring recalculations. This decreases the chip’s effective computational power, particularly for workloads involving conv operations. Therefore, aggressive level adjustment and $\beta$ selection involve a trade-off. For tasks that require faster response, such as real-time inference, enabling aggressive adjustment is not recommended. For edge computing devices requiring better energy efficiency, $\beta$ can be set as needed.

\begin{figure}[t]
    \centering
    \includegraphics[width=\linewidth]{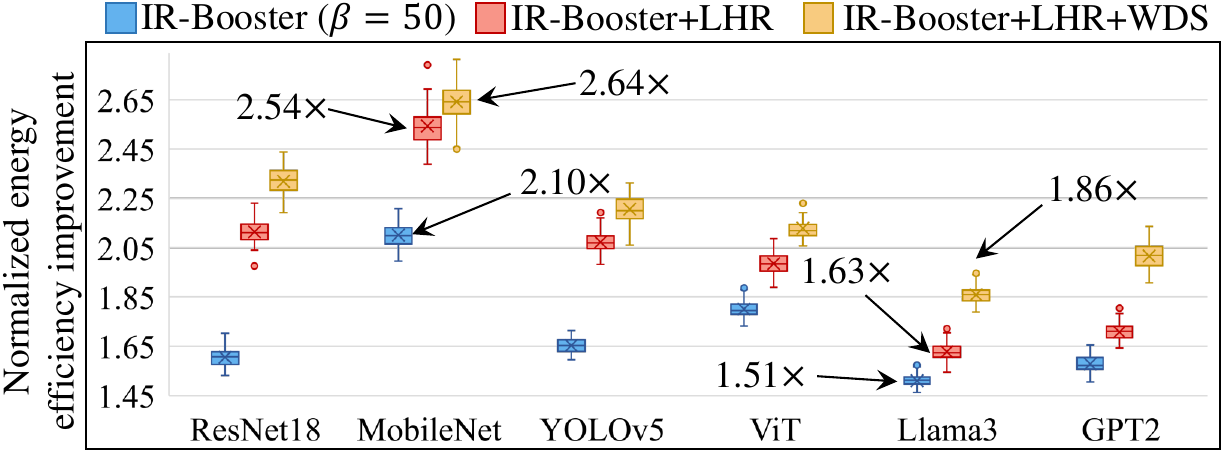}
    \caption{\revision{Energy efficiency enhancing of \booster\ and the improvement of HR optimization method (\lhr\ and \wds)}} 
    \label{fig:booster-energy}
\end{figure}

As shown in Figure~\ref{fig:booster-energy}, \booster\ alone provides significant energy efficiency improvements (1.51\thicktilde2.10$\times$), with \lhr\ and \wds\ providing even greater benefits.



\subsection{HR-aware Task Mapping Validation}
Traditional task mapping algorithms consider macros within a PIM chip to be equivalent and therefore rely on naive mapping methods, such as sequential mapping and zigzag mapping~\cite{tangram}.

\begin{figure}[t]
    \centering
    \includegraphics[width=0.95\linewidth]{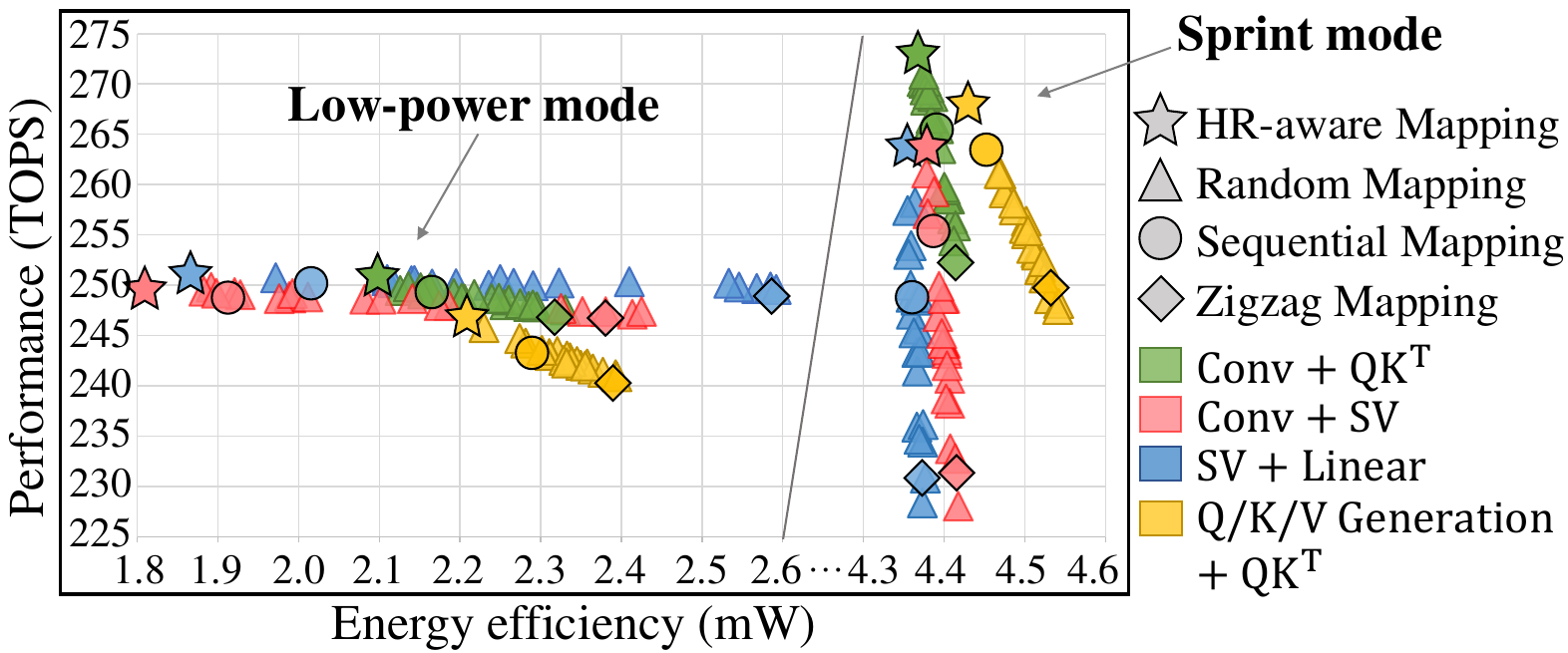}
    \caption{HR-aware task mapping v.s. others} 
    \label{fig:mapping}
\end{figure}

Figure~\ref{fig:mapping} compares the effects of \HR-aware task mapping and traditional mapping across different operator combinations. The results indicate that mapping methods that do not explicitly account for \HR\ result in suboptimal performance and energy efficiency under corresponding operating modes. In terms of energy efficiency, poor mapping schemes often group tasks with significantly different \HR\ levels together, which increases power consumption and reduces energy efficiency. In terms of performance, when tasks from different operators are assigned to the same group, a failure in any one task can suspend all tasks within that group. This suspension propagates delays to other tasks of the same operator, severely degrading the overall computational performance.

In contrast, \HR-aware task mapping explicitly incorporates \HR\ into the mapping process, fully leveraging the IR-drop mitigation benefits of \wds, \lhr, and \booster. This approach delivers significant improvements in both performance and power efficiency, underscoring its critical role in maximizing hardware potential.

\subsection{\revision{Overhead Analysis}} \label{subsec:overhead}
\subsubsection{\revision{Performance overhead}}\ \par
\revision{We have taken a series of measures to avoid the additional delay introduced by the \name\ method, such as introducing pipelines in the Shift Compensator to overlap delays and fine-grained macro pause control to reduce mutual interference. However, some overheads are inevitable since V-f adjustments are not instantaneous, and recalculations also introduce extra delay. As shown in Figure~\ref{fig:ablation}, an overly aggressive \booster\ strategy may lead to little performance loss for certain workloads compared to a conservative strategy. However, despite these unavoidable overheads, the various \name\ components still result in significant IR-drop mitigation, along with improvements in energy efficiency and overall performance.}

\subsubsection{Area and power overhead}\ \par
To ensure that the shift compensator (SC) does not impact macro throughput, we implement a pipelined design. This approach leverages the PIM macro’s inherent structure, allowing all banks (typically 32, 64, or even 128) within a macro to share a single compensator for lightweight correction computations. Thanks to its simple and efficient design, the compensator introduces minimal overhead, contributing less than 0.2\% to the overall chip area and less than 1\% to the power consumption of the PIM chip.

\booster\ adds overhead for IRFailure detection and V-f pair control. Fortunately, due to lower resolution requirements, the simplified IR Monitor design, based on~\cite{IR-Monitor}, incurs less than 0.1\% area overhead and less than 0.5\% additional power consumption according to synthesis results from~\cite{TOOL_hspice}. Moreover, the existing RISC-V core within the PIM chip—originally designed for macro control and general-purpose computations—can be extended to support V-f pair control operations. It ensures that the area and power overhead associated with control remains virtually negligible.

\section{Discussion and Future Work} \label{sec:discussion}

\begin{figure}[t]
    \centering
    \includegraphics[width=0.95\linewidth]{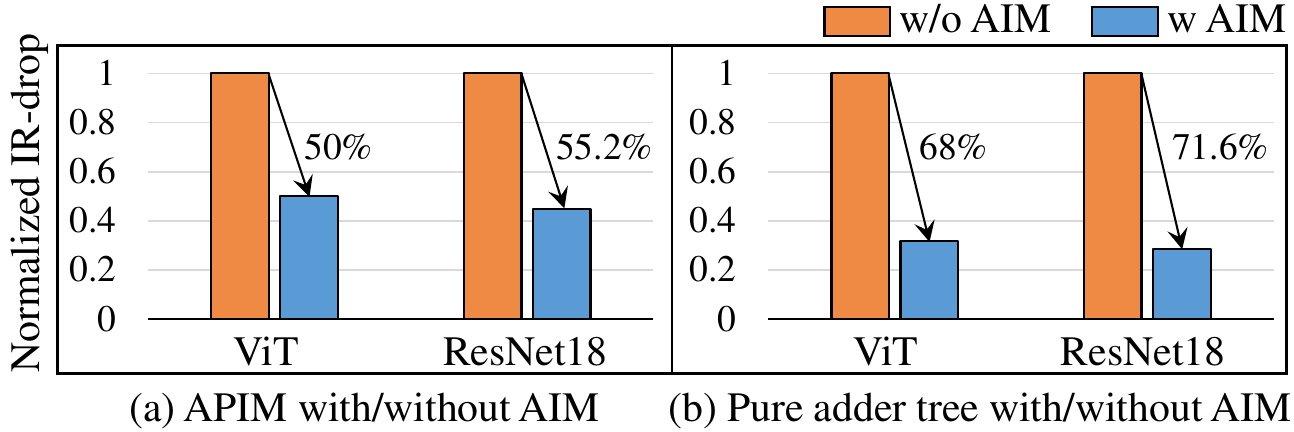}
    \caption{\name\ effects on APIM and pure adder tree} 
    \label{fig:acim and adder tree}
\end{figure}
We evaluate the impact of applying \name\ to a 28nm 128$\times$32 SRAM APIM macro using \cite{TOOL_redhawk-sem}. Figure~\ref{fig:acim and adder tree}-(a) illustrates that \name\ achieved approximately 50\% IR-drop mitigation on the APIM macro. This result is lower than the 58.5\%\thicktilde69.2\% achieved on DPIM. We attribute this disparity to the lower sensitivity of analog circuits to IR-drop mitigation. Additionally, we separately model and evaluate the bit-serial adder trees in the DPIM macro, which, as shown in Figure~\ref{fig:acim and adder tree}-(b), demonstrated notable IR-drop mitigation. These results highlight the potential of \name\ for mitigating IR-drop in PIM as well as other AI processors, such as TPUs and GPUs, which rely on complex digital multiplication and accumulation circuits.

\revision{\name\ can be extended to adapt to PIM designs that support floating-point data types \cite{DCIM3, DCIM4}, due to their bit-serial and in-situ calculation properties. While the design introduces mantissa alignment and exponent calculations, the core operations—MACs of mantissa—still rely on complement code integer multiplication, making the \lhr-like weight fine-tuning and \wds\ methods applicable. However, the relationship between \tog\ and IR-drop, as well as how these improvements affect the specific reduction in IR-drop, remains to be explored, and we leave this for future work.}

For TPUs and GPUs, a significant challenge arises from their non-bit-serial and non-in-situ MAC implementations, which complicates workload analysis and the correlation with IR-drop. Overcoming these challenges is a promising direction for future research.



\section{Conclusion} \label{sec:conclusion}

In this work, we propose \name\ from an architecture perspective to address the critical challenge of IR-drop in high-performance PIM, a growing threat to performance and reliability as circuit complexity and operating frequencies increase. By employing software and hardware co-design, \name\ effectively mitigates IR-drop while preserving computational accuracy, enhancing energy efficiency, and improving processing performance. It introduces architecture-level IR-drop metrics to quantify and analyze IR-drop characteristics, which form the foundation for proposed software optimizations aimed at IR-drop mitigation. Through the proposed \lhr\ and \wds\ methods, we effectively mitigate IR-drop via software optimization while computational accuracy is preserved. These software optimizations are further combined with hardware-based dynamic V-f pair adjustments for real-time adaptation by an HR-aware task mapping strategy to maximize overall improvement. Post-layout simulations on a 7nm 256-TOPS PIM design validate that \name\ achieves up to 69.2\% IR-drop mitigation, delivering a 2.29$\times$ improvement in energy efficiency and a 1.152$\times$ speedup, providing an efficient solution for IR-drop mitigation in high-performance PIM.

\begin{acks}
We appreciate the valuable feedback and constructive comments from all reviewers and our shepherd. This work is sponsored by National Natural Science Foundation of China (Grant No. 62032001), Beijing Natural Science Foundation L243001, and 111 Project (B18001).
\end{acks}

\bibliographystyle{ACM-Reference-Format}
\bibliography{ref}









\end{document}